%% file: anns.tex
\documentclass[10pt,sigplan,letterpaper,twocolumn]{article}
\usepackage[10pt,nocopyright]{sigmin}

\usepackage{listings}
\usepackage{color}
\usepackage{dingbat}
\usepackage[short]{optidef} 

\usepackage{fancybox}
\usepackage{epsfig,makecell}                                      
\usepackage[hyphens]{url}                                         
\usepackage[breaklinks,colorlinks]{hyperref}                      
\usepackage[usenames,dvipsnames,table]{xcolor}                    
\hypersetup{citecolor=blue,linkcolor=blue}                        
\usepackage{amsmath,amsopn,amssymb,amsthm}                        
\usepackage{thmtools}
\usepackage{thmbox}
\usepackage[toc,page]{appendix}
\usepackage{subcaption}                                           
\usepackage{soul}
\usepackage{endnotes,microtype,xspace,fancyvrb,multirow}          
\usepackage{epigraph}

\usepackage{booktabs}
\usepackage{tabularx}
\usepackage{array,underscore,relsize}                             
\usepackage[T1]{fontenc}                                          
\usepackage{times}                                                
\usepackage{fancyhdr,lastpage}                                    
\usepackage{enumitem}                                             
\usepackage[labelfont=bf,font=normal,skip=1pt]{caption}           
\usepackage[export]{adjustbox}                                    
\usepackage{breakurl}                                             
\usepackage{pifont}                                               
\usepackage{tablefootnote}                                        
\usepackage{bbding}                                               
\usepackage[table]{xcolor} 
\usepackage[linesnumbered,vlined,boxed,commentsnumbered,ruled]{algorithm2e}
\usepackage[normalem]{ulem}
\usepackage{float}
\DontPrintSemicolon

\usepackage{amssymb}
\usepackage[normalem]{ulem}
\usepackage[hang,flushmargin]{footmisc}
\usepackage{textcomp}
\usepackage{graphicx}
\usepackage{authblk}
\usepackage{stfloats}

\usepackage[compact]{titlesec}
\titleformat*{\section}{\large\bfseries}
\titleformat*{\subsection}{\normalsize\bfseries}
\titleformat*{\subsubsection}{\normalsize\bfseries}
\titlespacing{\section}{0pt}{3ex}{1ex}
\titlespacing{\subsection}{0pt}{2ex}{1ex}

\hypersetup{
  colorlinks,
  linkcolor={red!50!black},
  citecolor={blue!50!black},
  urlcolor={blue!80!black}
}

\newcommand{\stitle}[1]{\vspace{1.2ex}\noindent{\bf #1}\,}
\newcommand{\nospacestitle}[1]{\noindent{\bf #1}\,}
\newcommand{\sys}{\textsc{OMEGA}}
\newcommand{\etitle}[1]{\vspace{0.5ex}\noindent{\em \ul{#1}}}
\newcommand{\company}{\textsc{Alibaba}}

\newcommand{\laet}{{{LAET}}}
\newcommand{\darth}{{{DARTH}}}

\newcommand{\fig}[1]{Figure{~\ref{#1}}}

\newcommand{\one}{\texttt{\uppercase\expandafter{\romannumeral1}}}
\newcommand{\two}{\texttt{\uppercase\expandafter{\romannumeral2}}}

\newcommand{\pcomment}[2]{{\bf[\textcolor{red}{#1}: \textcolor{blue}{#2}]}}

\newcommand{\TODO}[1]{\textcolor{red}{TODO: #1}}

\newcommand{\FJF}[1]{{\pcomment{FJF}{#1}}}

\usepackage{tikz}
\usepackage{colortbl}
\usetikzlibrary{tikzmark,calc}

\usepackage{balance}
\newcolumntype{P}[1]{>{\centering\arraybackslash}p{#1}}

\definecolor{appcolor}{RGB}{191,255,255}

\SetKwInput{KwInput}{Input}
\SetKwInput{KwOutput}{Output}
\SetKwComment{Comment}{$\triangleright$\ }{}

\providecommand{\rmkhere}[2][none]{}

\begin{document}

\title{\Large \bf Efficient K-generalizable Learned Search}

\makeatletter
\renewcommand\AB@affilsepx{ \quad\protect\Affilfont \, } 
\makeatother

\setlength{\affilsep}{0.5em}
\author[1]{Yifan Peng}
\author[2]{Jiafei Fan}
\author[1]{Xingda Wei\,{\Envelope}}
\author[3]{Sijie Shen}
\author[1]{Rong Chen}
\author[3]{Jianning Wang}
\author[3]{Xiaojian Luo}
\author[3]{Wenyuan Yu}
\author[3]{Jingren Zhou}
\author[1]{Haibo Chen}

\affil[1]{\vspace{-2.mm}Shanghai Jiao Tong University}
\affil[2]{Boston University}
\affil[3]{Alibaba Group\vspace{-1.mm}}

\date{}
\maketitle
\def\thefootnote{\Envelope}\footnotetext{Xingda Wei is the corresponding author (\url{wxdwfc@sjtu.edu.cn}).}

\renewcommand{\thefootnote}{\arabic{footnote}}
\frenchspacing

\input{abs}

\input{introduction}

\input{bg}



\input{overview-v1}

\input{design-v1.tex}


\input{eval-v1.tex}


\input{relatedwork}

\input{conclusion}

\section*{Acknowledgments}
\input{ack}

\balance

\small{
\bibliographystyle{acm}
\bibliography{anns}
}


\twocolumn
\appendix
\input{appendix}

\clearpage

\end{document}

%% file: abs.tex
\begin{abstract}
\noindent
Learned top-$K$ search improves the accuracy--latency trade-off of
graph-based vector search, but existing methods target a fixed $K$. 
Serving real-world workloads with various $K$ values thus incurs preprocessing cost
proportional to the number of distinct $K$s.

We present {\sys}, which supports efficient arbitrary $K$ learned search with the preprocessing
cost of one top-$1$ model.
It achieves so by reducing learned top-$K$ search to repeated masked learned top-$1$ search with refinement,
using carefully crafted trajectory features that generalize across $K$s.
{\sys} further employs rank-wise confidence allocation to tame error accumulation
during refinements
and applies a statistical forecast method to reduce model invocation overhead
without compromising recall. 
Across nine dataset-scale configurations, {\sys} meets the 0.95 recall target
and reduces mean latency by up to 65\% (average: 23\%; median: 20\%)
compared with learned search methods (including LAET, DARTH,
MultiK-DARTH, ANNiE, and Ada-ef)
at their lowest-preprocessing configurations.
On production traces, it reduces total CPU time by up to 28\% versus
traditional HNSW search. 

\end{abstract}

%% file: introduction.tex
\section{Introduction}
\label{sec:intro}

\noindent
Vector databases have become a fundamental cloud service
powering \emph{latency-sensitive} applications---from personalized recommendation engines~\cite{DBLP:conf/middleware/LiLGCNWC18,DBLP:conf/kdd/LiLJLYZWM21,DBLP:journals/pvldb/WeiWWLZ0C20}
to retrieval-augmented generation (RAG)~\cite{DBLP:conf/nips/LewisPPPKGKLYR020,langchain}.
The key operation is approximate nearest neighbor search (ANNS):
given a query vector and $K$ results required,
the search finds the top-$K$ most similar vectors from a collection of vectors.
To accelerate ANNS, vectors are organized into indices,
where graph-based indices like HNSW~\cite{hnswlib} and NSG~\cite{NSG} are
the backbone of many production vector databases~\cite{milvus,faiss,pgvector}.
Both latency and accuracy are critical for serving vector queries,
e.g., the recommendation engine needs to return relevant results as soon as possible
while an incorrect result may mislead users.

Graph-based ANNS inherently faces an accuracy–latency trade-off,
because (1) the amount of search
required to retrieve the top-$K$ results varies widely across queries and is unknown a priori, 
and (2) due to the uncertainty,
current indices can only configure a fixed search budget that is applied uniformly to all queries with the same $K$.
As a result,
a large budget yields higher accuracy but incurs higher latency due to over-search for many queries,
whereas a small budget reduces latency at the cost of accuracy due to under-search.
A promising technique proposed recently is \emph{learned search}~\cite{laet,darth,annie} 
aiming to achieve accuracy and low latency simultaneously:
given a $K$, 
they train a model to predict whether the recall target has been met or how much search is needed to meet it.
The search can then stop before reaching a conservative fixed budget,
reducing latency while satisfying the accuracy requirement.

While state-of-the-art learned search methods such as DARTH~\cite{darth}, LAET~\cite{laet}, and ANNiE~\cite{annie} achieve both high accuracy and performance for a target $K$,
we still found a significant gap in real-world deployments.
First, real-world vector databases must serve queries with \emph{diverse and possibly changing} $K$ values (multi-$K$ queries):
$K$ is determined by downstream serving needs (e.g., multi-stage retrieval pipelines) rather than fixed by the index.
Meanwhile, vector databases already consume non-trivial preprocessing budgets
for background tasks such as index compaction (\textsection{\ref{sec:bg-motiv}}),
and because vendors absorb these costs rather than passing them to users,
any speedup method must keep its additional preprocessing overhead small.

The aforementioned features pose a key issue for learned search methods like DARTH~\cite{darth}, LAET~\cite{laet}, and ANNiE~\cite{annie}:
they all targeted a fixed-$K$ serving scenario, and generalizing them to multiple $K$ requires 
preprocessing costs proportional to the number of targeted $K$s.
In our measurement (detailed in \textsection{\ref{sec:overview}}), 
a straightforward approach of training models that cover 39\,\% of the most frequently accessed $K$s in a sampled cluster in {\company} 
leads to a 1.95\,$\times$ increase in total preprocessing cost (compared to 1.31\,$\times$ for a single $K$).
Such a high preprocessing cost is prohibitive for real-world deployments,
because the original preprocessing cost is already 22\,\% of the serving cost.

We present {\sys}, the first generalizable learned search framework for multi-$K$ graph-based ANNS with constant small preprocessing cost.
{\sys} requires only the training time of one top-$1$ model (i.e., a model trained on $K{=}1$), 
yet it can effectively accelerate arbitrary-$K$ queries with high accuracy.
The key idea is to reduce top-$K$ learned search to $K$ iterative top-$1$ learned searches:
after finding the current top-$r$ results ($r < K$), {\sys} masks them from the search state
and reuses the same top-$1$ model to find the $(r{+}1)$-th result.
This reduction is effective because we found that the search trajectory of finding the next nearest neighbor---regardless of how many have already been found---exhibits
a consistent pattern that a single model can learn (\textsection{\ref{sec:design-train}}).
This new iterative reduction is a fundamental innovation
that enables learned search to support real-world multi-$K$ searches
with low preprocessing budgets.\label{rev:iterative-reduction}
While this reduction eliminates per-$K$ retraining, it introduces two practical challenges.
First, finding $K$ results requires at least $K$ model invocations, and the accumulated cost
can offset the latency benefit of learned search for large $K$.
To address the first challenge, we identify a statistical property: given that the top-$N$ results have been found,
the probability of the current search set already containing the remaining top-$(K{-}N)$ results
can be accurately forecasted from a pre-profiled table, without invoking the model:
if the forecasted recall already meets the target, {\sys} stops early, skipping model calls for larger ranks (\textsection{\ref{sec:design-inference}}).
Second, prediction errors can accumulate across iterations, causing recall to degrade for large $K$.
To improve robustness, {\sys} applies a rank-wise confidence allocation scheme
that imposes stricter stopping thresholds on earlier iterations,
whose errors would otherwise propagate more widely.

{\sys} meets the 0.95 recall target across all nine datasets.
Compared with DARTH, MultiK-DARTH, and LAET at their lowest-preprocessing
configurations, it reduces mean latency by 7--36\%, 3--25\%, and 8--21\%,
respectively, on BIGANN, BIGANN-1B, DEEP, and three production workloads. On GIST,
Text2Image, and MS MARCO, its latency is within 9\% of DARTH and
MultiK-DARTH while retaining the advantage of one $K$-independent model.
On production traces, {\sys} further reduces total CPU time for serving and
preprocessing by up to 28\% compared with traditional HNSW search using a conservative fixed budget.

{\sys} is open-sourced at {\burl{https://github.com/driPyf/OMEGA}}.

\stitle{Discussion. \,}
{\sys}'s end-to-end benefit depends on the
compaction frequency,
e.g., less frequently updated datasets may benefit less from {\sys}.
Nevertheless, {\sys} is a general method
that lowers the training cost while improving multi-$K$ search performance
for graph-based vector search, and we have validated it on {\company}'s
representative workload, whose trace is drawn from an entire production cluster
serving many customers over 24 hours (\textsection{\ref{sec:motivation}}).

%% file: bg.tex
\section{Serving Vector Search in the Wild}
\label{sec:bg-motiv}

\subsection{Preliminaries and System Context}
\label{sec:bg}

\nospacestitle{Top-$K$ ANNS and its accuracy. \,}
Vector databases execute user queries through \emph{approximate nearest neighbor search} (ANNS):
given a query vector and a user-specified $K$, the system retrieves the $K$ most similar vectors from a vector collection,
where similarity is measured by metrics such as Euclidean distance or cosine similarity.
Because exact search is computationally infeasible at scale,
ANNS algorithms trade a small amount
of accuracy for orders-of-magnitude speedups.
Specifically, search accuracy is quantified by \emph{recall@K},
defined as $ \frac{|G \cap R|}{K} $,
where $R$ denotes the set of $K$ vectors returned by the approximate search,
and $G$ is the ground-truth top-$K$ nearest neighbors of the query.

\stitle{Graph indices for top-$K$ ANNS. \,}
Graph-based indices such as HNSW~\cite{hnswlib}, NSG~\cite{NSG},
and Vamana~\cite{diskann} are predominant in practice~\cite{hnswapp1,hnswapp2,hnswapp3},
with HNSW being the primary choice in {\company},
thanks to their superior recall-latency trade-offs and resource efficiency.
Although these variants differ in graph structure and search strategy,
they share the same methodology:
organizing vectors into a proximity graph where nearby vectors are connected by edges.
The search follows a \emph{best-first approach},
exploring the graph from an entry point
and progressively collecting candidates into a \emph{search\_set}.
The search depth (\emph{steps}) is controlled by a global hyperparameter (e.g., \emph{efSearch} in HNSW),
and once exhausted, the top-$K$ vectors in the search\_set are returned.

\begin{figure}[!t]
    \centering
    \includegraphics[width=\PaperLayout{0.92}{0.55}\linewidth]{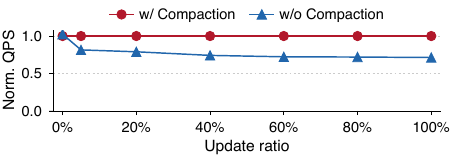}\PaperLayout{\\[0pt]}{\\[-7pt]}
    \begin{minipage}{1\linewidth}
    \caption{\small{The necessity of compaction under evolving workloads. }}
    \label{fig:compaction-qps-new}
    \end{minipage}\PaperLayout{\\[-15pt]}{\\[-12pt]}
\end{figure}

\stitle{Evolvable vector indices. \,}
Modern vector database workloads
involve continuous insertions and deletions,
which require updating the index accordingly.
Since incremental updates degrade graph quality,
a common practice is background compaction~\cite{proxima,guo2022manu,zhong2025vsag}:
updates are first buffered in a separate \emph{mutable} index,
and once the buffer exceeds a threshold,
a background job rebuilds the full index by merging the mutable index in.
Compaction is critical for both accuracy and performance,
as searches must otherwise query both indices.
{\fig{fig:compaction-qps-new}} illustrates this impact on throughput (queries per second, QPS)
for the DEEP~\cite{deep} benchmark.
Since DEEP is static, we emulate an evolving workload by building the base index on its first 1M vectors ($[0, 1M)$) and gradually inserting the next 1M ($[1M, 2M)$).
We use the same graph index as {\company}, whose internal index is HNSW,
together with the buffered updates described above.
Without compaction, QPS drops by 18.5--28.7\% as the fraction of updates in the workload grows,
confirming the necessity of compaction for evolving workloads.\label{rev:compaction-qps-new}

\begin{figure}[!t]
        \begin{minipage}{1\linewidth}
        \centering
        \includegraphics[width=\PaperLayout{0.95}{0.6}\linewidth, trim=0.25cm 20.7cm 21.0cm 0.25cm, clip]{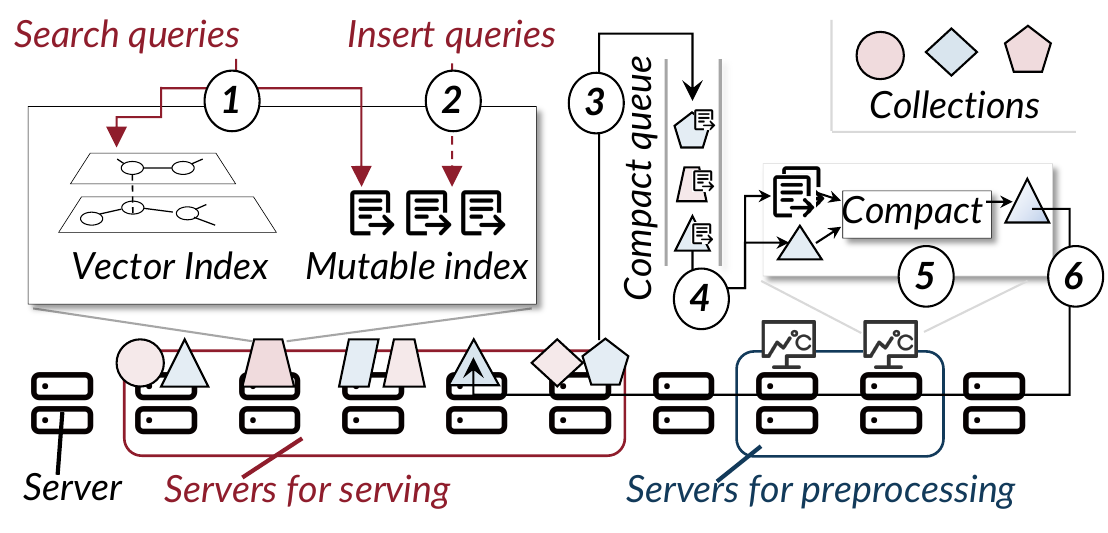}
        \end{minipage} \PaperLayout{\\[2pt]}{\\[-10pt]}
        \begin{minipage}{1\linewidth}
        \caption{\small{%
           Serving vector queries in the wild. 
        }}
        \label{fig:setup}
        \end{minipage} \PaperLayout{\\[-20pt]}{\\[-12pt]}
        \end{figure}

\stitle{Serving vector queries in the wild. \,}
{\fig{fig:setup}} illustrates the end-to-end serving architecture at {\company}.
{\company} hosts many applications,
each with an independent vector database called a \emph{collection}.
A collection maintains a graph index (typically HNSW) for searches (\ding{192})
and a mutable index for insertions (\ding{193}).
When compaction is needed, the collection ID is enqueued (\ding{194}),
and a dedicated fleet of servers fetches it (\ding{195}),
rebuilds the index (\ding{196}), and updates the collection (\ding{197}).

For usability, {\company} transparently sets hyperparameters for each collection,
most critically the maximum search steps that govern recall and latency.
These are tuned empirically, with conservatively large values to ensure high recall (detailed in \textsection{\ref{sec:current-works}}).
{\company} charges users for serving resources but not for the preprocessing fleet;
users can either provision servers explicitly or use a serverless pay-as-you-go model.

\begin{figure}[!t]
        \begin{minipage}{1\linewidth}
        \centering
        \includegraphics[width=\PaperLayout{1.0}{0.65}\linewidth,trim=0.25cm 24.2cm 23.85cm 0.25cm, clip]{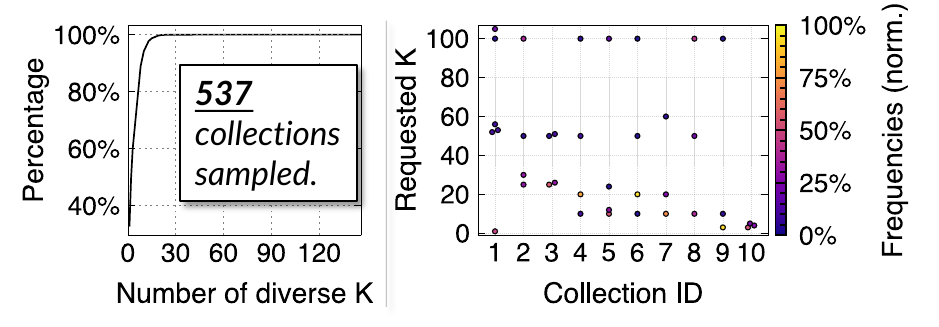}
        \end{minipage} \PaperLayout{\\[5pt]}{\\[-8pt]}
        \begin{minipage}{1\linewidth}
        \caption{\small{%
            (a) Distribution of the number of distinct $K$ values across sampled collections.
            (b) Breakdown of $K$ access patterns for sampled collections, with frequencies normalized (norm.) to total accesses per collection.
        }}
        \label{fig:multi-k}
        \end{minipage} \PaperLayout{\\[-10pt]}{\\[-8pt]}
        \end{figure}

\subsection{Characterizing Real-world Vector Searches}
\label{sec:motivation}

\nospacestitle{Beyond Single-$K$: Real-world workloads demand efficient multi-$K$ searches. \,}
Real-world applications issue queries with different $K$ values to each collection,
because $K$ is determined by downstream serving needs~\cite{DBLP:conf/recsys/CovingtonAS16,DBLP:conf/coling/LuHMN24,DBLP:conf/kdd/RenYWXWZ24} rather than fixed by the index.
For example, a recommendation app may first retrieve the top-10 results for one page
and then retrieve the top-20 when the user scrolls to the next.
{\fig{fig:multi-k}} (a) shows the distribution of distinct $K$ values
observed across 537 collections sampled from a production cluster at {\company} over 90 days:
56.1\% of the collections serve queries with more than two distinct $K$ values,
and 22.5\% serve more than three.
{\fig{fig:multi-k}} (b) further breaks down the per-collection $K$ distributions,
showing that some collections see uniform $K$ values while others are heavily skewed.

\begin{figure}[!t]
        \begin{minipage}{1\linewidth}
        \centering
        \includegraphics[width=\PaperLayout{0.93}{0.55}\linewidth]{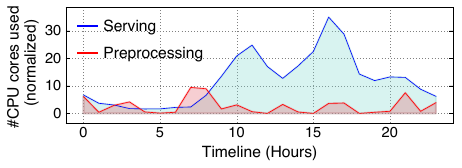}
        \end{minipage} \PaperLayout{\\[3pt]}{\\[-8pt]}
        \begin{minipage}{1\linewidth}
        \caption{\small{%
        A profile of the CPU computation power provisioned for serving
        and preprocessing vector collections in one cluster.
        }}
        \label{fig:preprocessing-costs}
        \end{minipage} \PaperLayout{\\[-15pt]}{\\[-16pt]}
        \end{figure}

\stitle{Beyond serving: Minimizing preprocessing costs is equally important. \,}
Since {\company} does not charge users for preprocessing,
the cost of index compaction falls entirely on the provider and is already significant.
{\fig{fig:preprocessing-costs}} shows the CPU cores dedicated to preprocessing versus serving in one of {\company}'s clusters:
preprocessing already consumes 22\,\% of the serving compute,
even though each collection is compacted only about once per day.
This is because (1) a single cluster hosts hundreds of collections---each compacted at least daily and more often during rapid evolution---so the costs accumulate;
and (2) each compaction averages 132 CPU core-minutes, equivalent to serving 792,000 queries.
Therefore, for multi-$K$ learned search, reducing serving latency alone is insufficient; the offline cost of supporting multiple $K$ values must also be controlled.

\subsection{Issues of Existing Approaches}
\label{sec:current-works}

\begin{figure}[!t]
        \begin{minipage}{1\linewidth}
        \centering
         \includegraphics[width=\PaperLayout{0.92}{0.6}\linewidth, trim=0.25cm 22.9cm 23.85cm 0.25cm, clip]{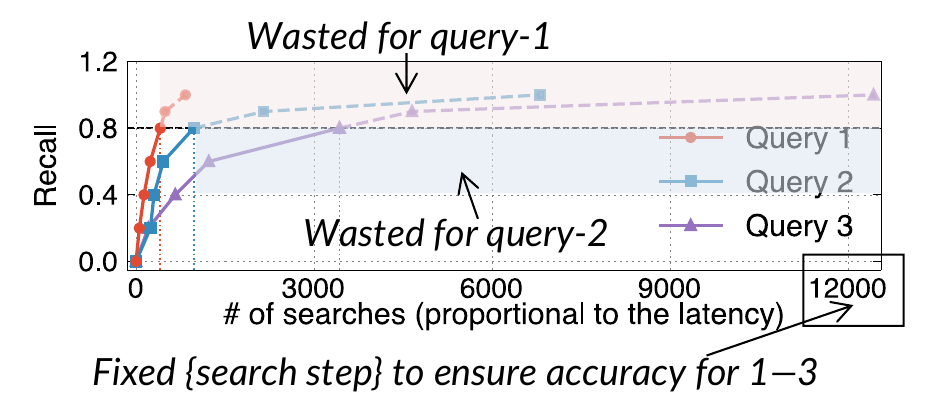}
        \end{minipage} \PaperLayout{\\[5pt]}{\\[-7pt]}
        \begin{minipage}{1\linewidth}
        \caption{\small{%
Performance (search latency) vs.\ accuracy (recall) trade-off for three sampled queries from a production collection at {\company}.
        }}
        \label{data:query-difficulty}
        \end{minipage} \PaperLayout{\\[-10pt]}{\\[-8pt]}
        \end{figure}     

\begin{figure*}[!t]
        \hspace{-2mm}
        \begin{minipage}{1\linewidth}
        \centering
        \PaperLayout{
                \includegraphics[width=\textwidth, trim=0.25cm 19.1cm 27.65cm 0.25cm, clip]{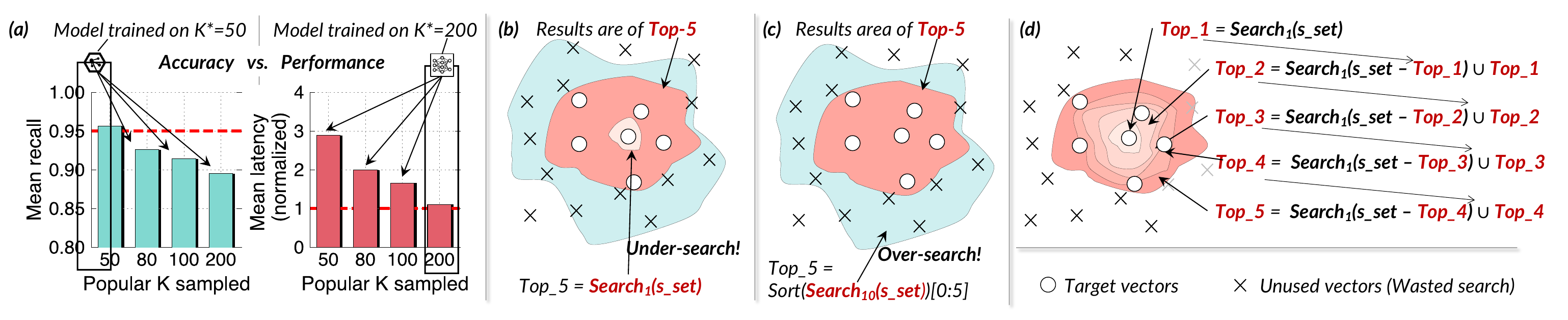}
        }{
                \includegraphics[width=0.7\textwidth, trim=0.25cm 9cm 53cm 0.25cm, clip]{figures/overview-camera.pdf}
        }
        \end{minipage} \PaperLayout{\\[-9pt]}{\\[-5pt]}
        \begin{minipage}{1\linewidth}
        \caption{\small{%
        (a) A model trained on one $K$ fails to generalize to other $K$s.
        (b) Under-searching when generalizing from small to large $K$.
        (c) Over-searching when generalizing from large to small $K$.
        (d) Our approach: a $K$-generalizable learned search using a single top-$1$ model ($Search_1$)
        that is reused iteratively on a dynamically evolving search set (\texttt{s\_set}).
        }}
        \label{fig:overview}
        \end{minipage} \PaperLayout{\\[-10pt]}{\\[-17pt]}
        \end{figure*}

\nospacestitle{Searches with fixed (heuristic-based) hyperparameters face a trade-off between
performance and accuracy. \,}
Existing open-source and commercial vector databases~\cite{faiss,hnswlib,milvus,weaviate,pgvector}
use one global setting for search steps across all queries in a collection,
leading to a well-known trade-off between performance and accuracy---either
missing the accuracy target or incurring high latency for some queries~\cite{darth,10.1145/3711896.3737383,DBLP:journals/pvldb/WangWCWPW24,DBLP:journals/vldb/ZoumpatianosLIP18}.
The root cause is that the searched candidate set size varies across queries
(even for the same $K$) and is unknown a priori, as shown in {\fig{data:query-difficulty}}.
For example, setting the search steps to 12,000 to satisfy Query~3
would significantly slow down Queries~1 and~2, which need far fewer search steps.

\stitle{Existing learned search is only effective on queries with one $K$. \,}
To break this trade-off,
recent works train a model to predict when to stop searching,
avoiding unnecessary search steps.
Given a query and its input features,
these models either predict the number of steps to search~\cite{laet}
or predict the current recall~\cite{darth}.
Taking DARTH~\cite{darth}---the state-of-the-art and most accurate
learned method (see \textsection{\ref{sec:eval}})---as an example:
once the predicted recall reaches the target (e.g., 0.95),
the search stops early.
DARTH achieves significant performance improvements while maintaining the same recall.

\begin{figure}[!t]
        \begin{minipage}{1\linewidth}
        \centering
        \includegraphics[width=\PaperLayout{0.99}{0.70}\linewidth]{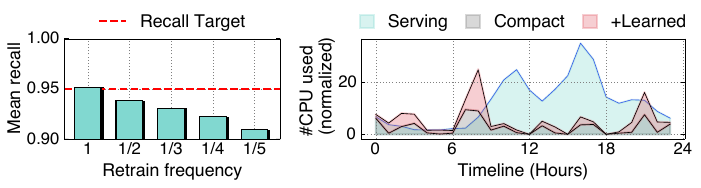}
        \end{minipage} \PaperLayout{\\[3pt]}{\\[-7pt]}
        \begin{minipage}{1\linewidth}
        \caption{\small{%
        (a)  The necessity of retraining after index compaction to maintain high recall.
        (b)  The additional preprocessing costs introduced by the state-of-the-art learned approach DARTH~\protect\cite{darth}.
        }}
        \label{fig:data-existing-solutions}
        \end{minipage} \PaperLayout{\\[-10pt]}{\\[-15pt]}
        \end{figure}

However, this fixed-$K$ training cannot serve multiple $K$ values.
As shown in {\fig{fig:overview}} (a),
when a model trained for one $K^*$ is used for another $K$,
it either loses accuracy (left) or loses performance (right).
For $K > K^*$, the model under-searches,
because the search area for $K^*$ is smaller than that for larger $K$ values ({\fig{fig:overview}} (b)).
For $K < K^*$, the model over-searches,
exploring an unnecessarily large area and increasing latency ({\fig{fig:overview}} (c)).

\vspace{1.5mm}
\noindent
\shadowbox{%
  \begin{minipage}{0.96\linewidth}
    \noindent
        Learned searches can achieve both high accuracy and performance on a specific $K$.
        However, real workloads serve multiple $K$ values, and models trained for one $K$ are ineffective on others.
        We need a learned search that generalizes across $K$ values with minimal preprocessing cost.
  \end{minipage}%
}\\[-20pt]

%% file: overview-v1.tex
\section{Generalizable Learned Search}
\label{sec:overview}

\subsection{Generalizable Learned Search}
\label{sec:overview-generalizable}

\nospacestitle{Design goals of {\sys}. \,}
A $K$-generalizable learned search must satisfy three goals:
(1) \emph{Accuracy}: recall meets the application's target;
(2) \emph{Efficiency}: low search latency;
and (3) \emph{Low preprocessing cost}: training overhead remains small relative to compaction.

\stitle{Existing $K$-generalizable learned search and our approach. \,}
Achieving all three goals simultaneously is non-trivial: straightforward extensions of existing fixed-$K$ methods sacrifice accuracy, efficiency, or low preprocessing cost.
We examine these extensions below and then present our approach.

\vspace{3pt}\etitle{Train multiple models, each for a specific $K$. \,}
The most direct extension is to train one model per $K$ and dispatch accordingly at serving time.
This achieves (1)\,+\,(2) but not (3).
On a real-world trace from {\company}, training models for just 39\,\% of the most frequent $K$s
already increases preprocessing cost by 1.95\,$\times$ ({\fig{fig:data-existing-solutions}} (b)).
The cost is high because
(1) each compaction requires retraining to maintain accuracy ({\fig{fig:data-existing-solutions}} (a)),
and (2) reaching low training loss still requires a non-trivial training budget
({\fig{data:training-loss}}); in our measured setting, even after our
optimizations (\textsection{\ref{sec:design-train}}, \textsection{\ref{sec:eval-setup}}),
training one model already takes $\ge$\,32\% of compaction time.

GPU acceleration does not help:
on BIGANN-100M, replacing a 32-core CPU with an NVIDIA Tesla V100 reduces training time by only 5.1\,\%,
because the lightweight models used for low inference overhead~\cite{darth} are inherently hard to parallelize on GPUs
(Appendix~\ref{sec:appendix-gpu})---let alone the additional deployment cost of GPUs.

\vspace{3pt}\etitle{Train a model that generalizes to different $K$ values. \,}
Another approach is to train a single model that takes $K$ as an input feature,
using training samples from multiple $K$s.
This avoids maintaining multiple models.
Although it trains one model, its training set must 
cover every target $K$ represented in the workload,
and each $K$ requires enough samples to achieve high recall.
As a result, covering each additional $K$ still costs the training time of one $K$,
so the preprocessing time stays proportional to the number of $K$s in the workload.
We evaluate this approach by extending DARTH~\cite{darth} (\textsection{\ref{sec:eval}}).

\vspace{4pt}\etitle{Our approach: reduce top-$K$ to $K$ iterative top-$1$ searches. \,}
Our approach is to train only one top-$1$ model that predicts whether the current search set
contains the nearest vector, thereby minimizing preprocessing cost.
We show that this single model is sufficient for arbitrary top-$K$
by designing a mask-and-reuse procedure that reduces top-$K$ search to $K$ iterative top-$1$ searches.
{\fig{fig:overview}} (d) illustrates:
after finding the top-$1$ vector, {\sys} masks it from the search state
and reuses the same model to find the $2^{nd}$ nearest vector as a new top-$1$ problem.
The procedure iterates until all top-$K$ vectors are found.

\subsection{Challenges, approaches and {\sys} architecture}
\label{sec:overview-challenge}

\nospacestitle{C\#1. Ensure high recall. \,}
First, the learned model must generalize to masked search states. 
Naively reusing an existing model fails because its learned features
do not capture the essential structure under masking---e.g.,
the \emph{minimal distance} feature in DARTH~\cite{darth} cannot capture
the search dynamics toward the next top-1, leading to degraded recall. 
We propose \emph{distance-trajectory} features (\textsection{\ref{sec:design-train}})
that capture the distance variation trend during search
and are naturally robust to masking.
Second, errors accumulate across search iterations especially with larger $K$s. 
We address this with a rank-wise confidence allocation scheme
(\textsection{\ref{sec:design-robustness}})
that assigns stricter thresholds to early searches,
whose errors would otherwise propagate most widely.

\stitle{C\#2. Ensure high efficiency. \,}
Model invocation cost is a determining factor of learned search efficiency.
Our reduction of top-$K$ to $K$ iterative top-$1$ searches achieves generalizability with low preprocessing cost,
but amplifies this cost by up to $K{\times}$,
which is prohibitive for large $K$.
We observe that, as the search progresses and more top-$N$ ($N < r$) vectors are found,
the probability that the $r^{th}$ nearest vector is already in the search set
follows a predictable distribution that can be profiled offline.
By querying these offline-profiled distributions with each query's own search state
via lightweight table lookups,
we can forecast whether the search set already contains enough results
and skip model invocations entirely for later ranks
(\textsection{\ref{sec:design-inference}}).

\begin{figure}[!t]
        \begin{minipage}{1\linewidth}%
        \centering    
        \includegraphics[width=\PaperLayout{0.95}{0.7}\linewidth, trim=0.25cm 19.4cm 20.9cm 0.25cm, clip]{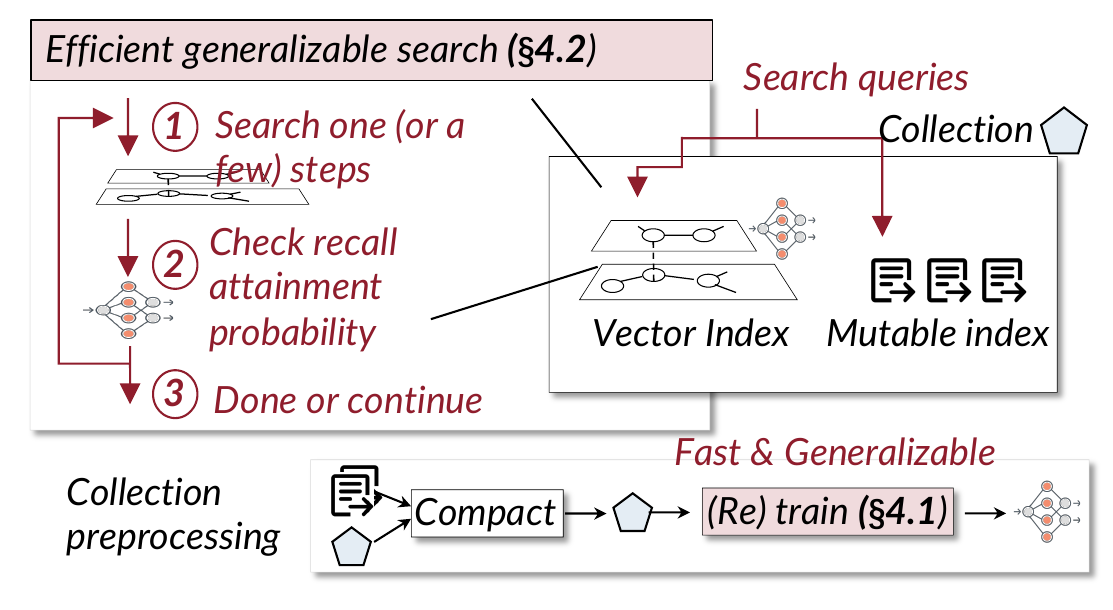}%
        \end{minipage} \PaperLayout{\\[3pt]}{\\[-10pt]}%
        \begin{minipage}{1\linewidth}%
        \caption{\small{%
            {\sys} architecture and basic execution flow. 
        }}
        \label{fig:arch}
        \end{minipage}\PaperLayout{\vspace{-15pt}}{\vspace{-8pt}}%
        \end{figure} 

\stitle{System architecture and serving flow. \,}
{\fig{fig:arch}} illustrates the architecture of {\sys}.
\emph{Offline}, after compaction builds the index,
a training pipeline produces a model
capable of predicting whether the current search set contains the top-$1$ vector
(\textsection{\ref{sec:design-train}}).
\emph{Online}, the model complements the existing vector index search:
during a single continuous search process,
we periodically evaluate the current search set to determine
whether it contains the top-$1$, top-$2$, $\ldots$, top-$K$ vectors iteratively,
stopping early once the statistical confidence of finding all top-$K$ vectors is sufficiently high
(\textsection{\ref{sec:design-inference}}).

%% file: design-v1.tex
\section{Detailed Design}
\label{sec:design}

\nospacestitle{Roadmap. \,}
Algorithm~\ref{alg:naive} gives the basic loop.
\textsection\ref{sec:design-train} details the model,
whose \emph{trajectory}-based features make it generalizable across ranks.
Two key refinements for large $K$ follow:
\textsection\ref{sec:design-robustness}
caps the compounding error across $K$ ranks with a non-uniform per-rank recall budget;
\textsection\ref{sec:design-inference} removes the $\Omega(K)$ model-invocation floor
via a statistics-based forecast.

\begin{algorithm}[t]
\caption{Basic {\sys} Search}
\label{alg:naive}
\DontPrintSemicolon
\KwIn{Graph index $G$, query vector $q$, recall target $r_{t}$, number of results $K$, {\sys} model $\mathcal{M}$}
\KwOut{$K$ nearest vectors of $q$}

$N \gets 0$ \tcp{Searched rank} 
$\mbox{search\_set} \gets \mathrm{PriorityQueue}()$\;
$\mbox{feature\_set} \gets \mathrm{FeatureSet()}$  \tcp{In \textsection{\ref{sec:design-train}}}

\While{$N < K$}{
    $\mbox{feature\_set}.\mathrm{mask}(\mbox{search\_set}[1..N])$\;
    \While{$\mathcal{M}.\mathrm{predict\_top\_1}(\mathrm{feature\_set}) < r_{t}$}{
        $\mbox{search\_set} \gets G.\mathrm{search\_one\_step}(q, \mbox{search\_set})$\;
        $\mbox{feature\_set}.\mathrm{update}(\mbox{search\_set})$\;
    }
    
    $N \gets N + 1$\;
}
\KwRet{$\mbox{search\_set}[1..K]$}\;
\end{algorithm}

\stitle{Basic search: top-$K$ with a top-$1$ model. \,}
Algorithm~\ref{alg:naive} describes the basic execution flow of our generalizable learned search method
assuming that we have a learned model trained
on a graph index (e.g., HNSW~\cite{hnswlib} or Vamana~\cite{diskann}) that can predict whether the current
search set already contains the target top-$1$ vector,
given the search set and the features collected during the search (\textsection{\ref{sec:design-train}}).
For a query $q$ and a recall target $r_{t}$ for a top-$K$ search,
we sequentially find the top-1, top-2, \dots, top-$K$ vectors
by continuously refining the search set (line 4--9).
For each rank $N$ (the search for the $(N{+}1)$-th nearest neighbor),
we first mask the previously found top-$N$ vectors in the feature set (line 5),
and then invoke $\mathcal{M}$ to estimate the probability that the current
search set already contains the (masked) top-$1$.
While this estimated probability is below $r_t$ (line 6),
we extend the search set by one more graph step (line 7) and update the features (line 8).
Once the estimate reaches $r_t$, we accept the current best unmasked candidate
as the rank-$(N{+}1)$ result and move on to the next rank.
After all $K$ ranks are accepted, the top-$K$ vectors in the search set are returned.

\subsection{Learning refinement with trajectory-based features}
\label{sec:design-train}

\begin{figure}[!t]
        \begin{minipage}{1\linewidth}
        \centering
         \includegraphics[width=\PaperLayout{0.95}{0.75}\linewidth, trim=0.25cm 15.9cm 16.4cm 0.05cm, clip]{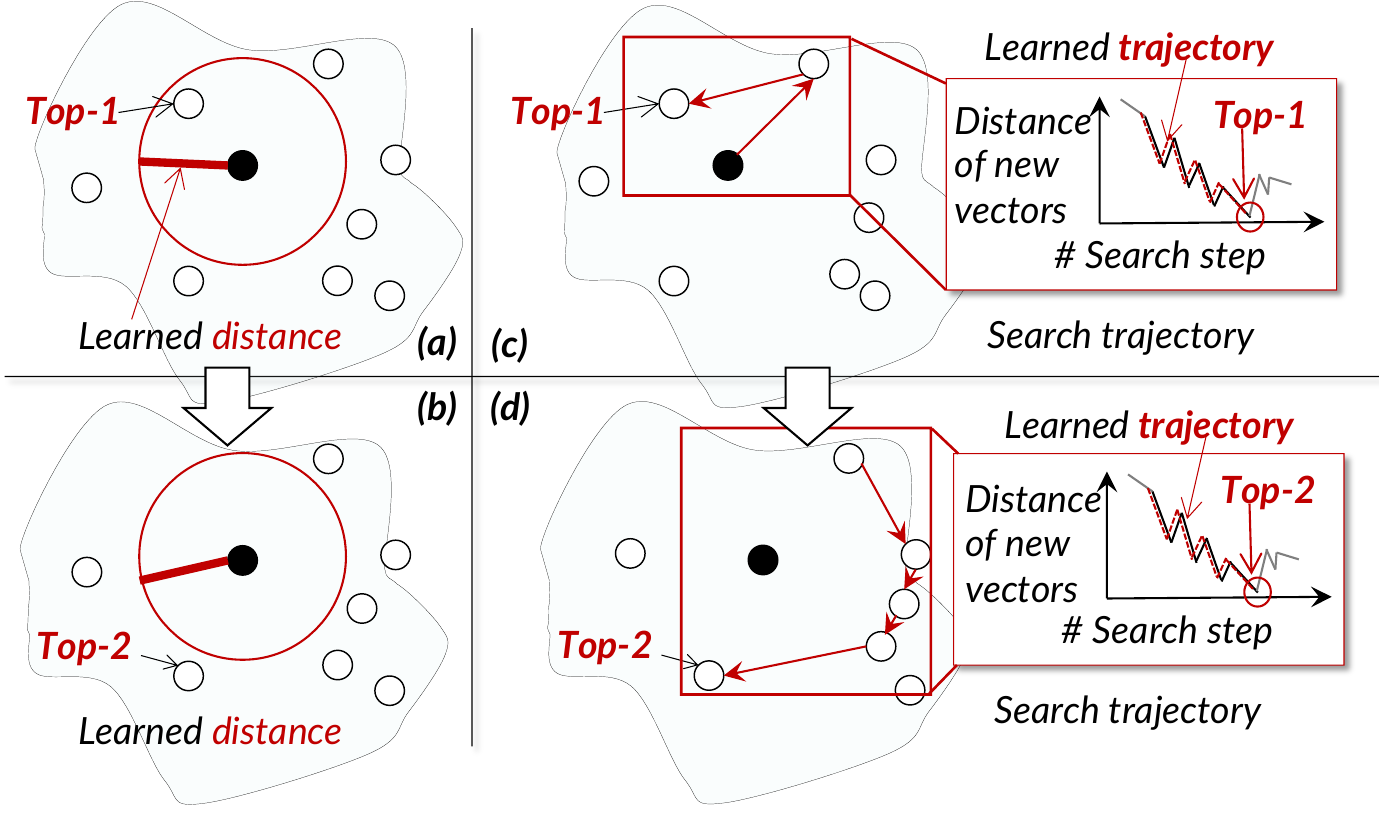}
        \end{minipage} \PaperLayout{\\[3pt]}{\\[-7pt]}
        \begin{minipage}{1\linewidth}
        \caption{\small{%
            (a,b)~DARTH's~\protect\cite{darth} absolute-distance features shift when vectors are masked;
            (c,d)~our trajectory-based features remain consistent.
            {\LARGE{$\bullet$}}~denotes the query vector.
        }}
        \label{fig:motiv-generalize-dist}
        \end{minipage} \PaperLayout{\\[-10pt]}{\\[-8pt]}
        \end{figure}

\noindent        
This section describes how we train an effective model that can
predict whether the current search set contains the target top-1 vector, 
even when some searched vectors are masked during refinement.

\begin{figure}[!t]
        \PaperLayout{\hspace{-3mm}}{}
        \begin{minipage}{1\linewidth}
        \centering    
        \includegraphics[width=\PaperLayout{1}{0.75}\linewidth, trim=0.25cm 20.1cm 12.2cm 0.25cm, clip]{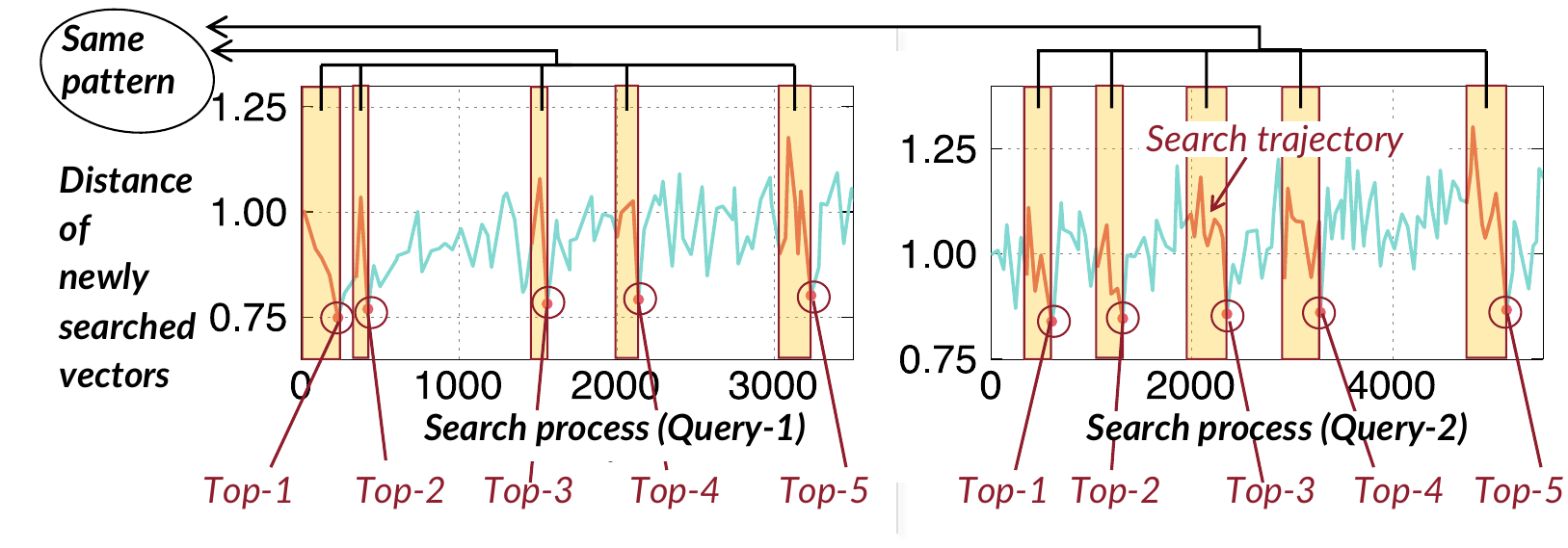}
        \end{minipage} \PaperLayout{\\[3pt]}{\\[-7pt]}
        \begin{minipage}{1\linewidth}
        \caption{\small{%
    Distance trajectories on HNSW~\protect\cite{hnswlib} for two queries on a production collection in {\company}.
    Each rank shows a sharply decreasing trend (yellow boxes); the pattern persists after earlier ranks are masked.
        }}
        \label{fig:search-trajectory-breakdown}
        \end{minipage} \PaperLayout{\\[-10pt]}{\\[-12pt]}
        \end{figure}

\stitle{Train the model with trajectory-based input features. \,}
Our key observation is that the \emph{distance trajectory}---the sequence of
distances between the query and each newly visited vector during the graph
search---exhibits a consistent, sharply decreasing trend as the search
approaches the $x^{th}$ nearest neighbor, regardless of rank $x$.
{\fig{fig:search-trajectory-breakdown}} illustrates this pattern
for two queries on a production collection;
we observe the same trend across all our evaluated datasets.
The pattern is not specific to HNSW:
we observe the same trend on Vamana~\cite{diskann}---another popular graph
index that underlies DiskANN (see Appendix~\ref{sec:appendix-features}).

This rank-invariance is precisely what prior features lack.
DARTH~\cite{darth} uses the absolute distances in the current search set
as input features;
as shown in {\fig{fig:motiv-generalize-dist}}\,(a,b),
these distances shift as $K$ changes because each rank has a different
absolute distance to the query,
so a model trained on top-1 does not transfer to other ranks.
In contrast, our trajectory captures the \emph{relative trend} of distances
during the search, which remains stable across ranks
({\fig{fig:motiv-generalize-dist}}\,(c,d)).

\begin{figure}[!t]
        \begin{minipage}{1\linewidth}
        \centering
         \includegraphics[width=\PaperLayout{0.95}{0.70}\linewidth]{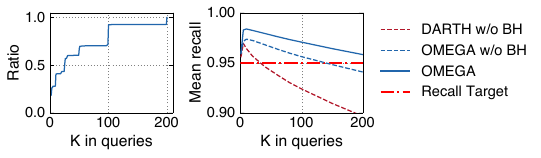}
        \end{minipage} \PaperLayout{\\[3pt]}{\\[-8pt]}
        \begin{minipage}{1\linewidth}
        \caption{\small{%
            (a)~Distribution of $K$ values from a production cluster in {\company}.
            (b)~Recall on DEEP-100M under Algorithm~\ref{alg:naive}: DARTH's
            absolute-distance features vs.\ {\sys}'s trajectory-based features,
            together with the rank-wise target allocation ablation of
            \textsection\ref{sec:design-robustness} (BH).
        }}
        \label{data:darth-vs-omega-on-call}
        \end{minipage} \PaperLayout{\\[-10pt]}{\\[-12pt]}
        \end{figure}

{\fig{data:darth-vs-omega-on-call}}\,(b) shows the effect of features
by running Algorithm~\ref{alg:naive} on DEEP-100M with DARTH's
absolute-distance features and {\sys}'s trajectory-based features ({\sys} w/o BH) as the
only difference in the base top-1 model.
The variant using DARTH's absolute-distance features fails to maintain high recall as $K$ grows,
while trajectory features alone sustain high recall across nearly the full
$K$ range observed in a production cluster at {\company}
({\fig{data:darth-vs-omega-on-call}}\,(a)).
Note that a small residual gap remains at the largest $K$, which stems from error
propagation across masked ranks rather than from the base model itself;
\textsection\ref{sec:design-robustness} closes this gap with a rank-wise
target allocation orthogonal to the choice of features ({\sys}).
\textsection{\ref{sec:ablation-study}} reports the full ablation on
additional datasets.

\stitle{Model choice and the trajectory window features ($w$). \,}
We adopt Gradient Boosting Decision Tree (GBDT)~\cite{friedman2001greedy,friedman2002stochastic,natekin2013gradient},
which our experiments confirm achieves strong accuracy on single-$K$ queries ({\fig{data:darth-vs-omega-on-call}}).
GBDT is also cheap to serve: $8$\,$\mu$s per invocation on average.
Note that the key is not only the model choice,
but the new feature proposed by us (see {\fig{fig:motiv-generalize-dist}}).

A challenge of adopting GBDT is that it requires fixed-length inputs.
Unlike prior work that adopted GBDT, such as DARTH,
our trajectory feature grows during the search. 
Therefore, 
we apply a sliding window of size $w$ over the most recent distances as the model's input.
We empirically set $w{=}100$, because we found accuracy is stable for $w \geq 50$ (\textsection{\ref{sec:sensitivity-to-hyperparameters}}).
As shown in {\fig{fig:search-trajectory-breakdown}},
even a short recent segment captures the decreasing trend. 

\begin{figure}[!t]
        \begin{minipage}{0.99\linewidth}
        \centering
         \includegraphics[width=\PaperLayout{0.99}{0.70}\linewidth,trim=0.25cm 25.2cm 23.75cm 0.25cm, clip]{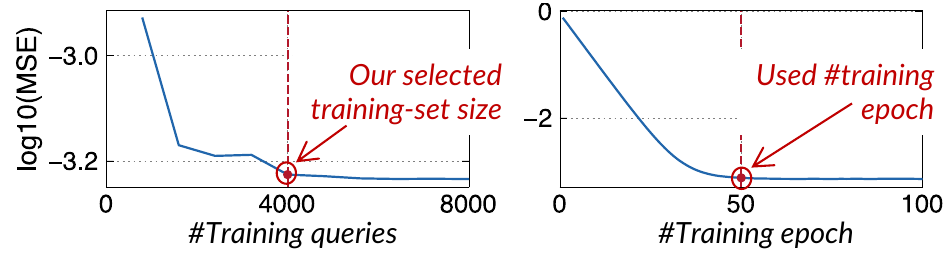}
        \end{minipage} \PaperLayout{\\[3pt]}{\\[-10pt]}
        \begin{minipage}{1\linewidth}
        \caption{\small{%
            Training loss (MSE) decreases as the number of training queries (left) or training epochs (right) grows.
        }}
        \label{data:training-loss}
        \end{minipage} \PaperLayout{\\[-15pt]}{\\[-8pt]}
        \end{figure}

\stitle{Training data and procedure. \,}
After a collection is constructed or compacted,
we train a GBDT with LightGBM~\cite{DBLP:conf/nips/KeMFWCMYL17}.
Each training record uses our trajectory features plus 
known effective features used in prior work,
i.e., \texttt{curr\_hops}, \texttt{curr\_cmps}, \texttt{dist\_1st}, and
\texttt{dist\_start}---the current search depth, cumulative distance
computations, the current top-$1$ distance, and the distance from the query
to the entry point, respectively; the label is a binary indicator of whether
the current search set already contains the target top-$1$ vector.

Two knobs jointly set the trade-off between preprocessing cost
and model accuracy: 

\emph{Training set.}
We use at least 4{,}000 queries---the minimum for high accuracy
across all our collections ({\fig{data:training-loss}}\,(a)).
When a collection lacks a training set,
we sample queries and label them by deep exploration on the HNSW index,
which costs only 3--15\,\% of the total
preparation time and is also required by prior
learned methods~\cite{laet,darth}.

\emph{Training epochs.}
Unlike DARTH's fixed 100 epochs~\cite{darth},
we early-stop once the loss plateaus
(e.g., epoch 50 in {\fig{data:training-loss}}\,(b)),
cutting preprocessing cost.
We also apply this early-stopping to DARTH for a fair comparison
in \textsection{\ref{sec:eval-setup}}.

\subsection{Search refinement with rank-wise confidence allocation}
\label{sec:design-robustness}

\nospacestitle{The issue and the solution. \,}
A key challenge of iterative top-$1$-to-top-$K$ generalization is that a single
incorrect early decision may corrupt subsequent masked searches.
Specifically, at each rank $i$, we accept the $i$-th neighbor once the top-$1$ model's confidence
exceeds the recall target $r_t$ (line 6 in Algorithm~\ref{alg:naive}),
with the assumption that the ground-truth top-$(i-1)$ neighbors have been found.
However, these neighbors are found using learned searches with possible errors. 
If each decision has a per-rank error rate $\epsilon = 1 - r_t$,
a union bound shows that the probability of at least one of the $K$ decisions being wrong
is at most $K\epsilon$.
Using $r_t$ as a uniform recall target across ranks
is therefore insufficient when scaling the $K$.

To cope with this, we observe that each masked decision is effectively one
hypothesis test---``has the $i$-th true neighbor already been found?''
The classical analogue
is multiple-testing with non-uniform costs: earlier tests are more consequential
because their errors propagate further. We therefore allocate a \emph{non-uniform}
recall budget across ranks, inspired by the step-up weighting of
Benjamini--Hochberg (BH)~\cite{Benjamini-Hochberg}: earlier ranks clear a tighter bar,
later ranks a looser one, and the resulting search empirically maintains the user target $r_t$.

\label{sec:design-bh-target}
Concretely, {\sys} sets $r_i = 1 - (1-r_t)\beta_i$ with
\[
\beta_i =
\frac{\sum_{j=1}^{i} 1/\sqrt{j}}
{\sum_{j=1}^{K} 1/\sqrt{j}} \in [0, 1].
\]
$\beta_i$ grows with $i$, so $r_1$ is close to $1$ (strictest) and $r_K = r_t$
(matches the user target). To keep the strictest ranks achievable by the base
model, we further cap $r_i \leftarrow \min(r_{\max}, r_i)$ with
$r_{\max} = (1+r_t)/2$---halfway between $r_t$ and $1$, which we empirically
find is reached by the base model on all evaluated datasets
(\textsection\ref{sec:sensitivity-to-hyperparameters}); values closer to $1$
make the inner loop fail to terminate on the hardest queries.

\stitle{Discussion on the theoretical analysis. \,}
We found that a rigorous theoretical analysis of the end-to-end
recall guarantee is difficult, even if we assume that the model can
guarantee correctness with a high, known probability
(as in the union-bound argument above).
This is because 
the \emph{cost} of an error at any rank is data-dependent: a wrongly accepted
vector often still belongs to the true top-$K$, so worst-case accounting
(charging all subsequent ranks to one early error) is overly conservative,
while average-case accounting would require distributional assumptions on
the index geometry.
Nevertheless, our empirical analysis in \textsection\ref{sec:ablation-study} 
shows consistent improvements with the search refinement.

\stitle{Improved search. \,}
Algorithm~\ref{alg:optimized-early-stop} shows the retrofitted search based on
Algorithm~\ref{alg:naive}: the changes with BH are that (1) at line 4 we precompute
$\{r_i\}_{i=1}^{K}$ once at the beginning of the search, and (2) at line 10 we replace the
global threshold $r_t$ in the inner stopping test with the rank-specific $r_{N+1}$.
Note that the amortized computation overhead of these weights is negligible in practice,
because they depend only on $K$,
so we precompute and cache them for each $K$ on the index.

\begin{algorithm}[t]
\caption{Optimized {\sys} Search}
\label{alg:optimized-early-stop}
\DontPrintSemicolon
\KwIn{Graph index $G$, query vector $q$, recall target $r_{t}$, number of results $K$, {\sys} model $\mathcal{M}$, 
profiled probability table $T_{prob}$}
\KwOut{$K$ nearest vectors of $q$}

$N \gets 0$ \tcp{Searched rank} 
$\mbox{search\_set} \gets \mathrm{PriorityQueue}()$\;
$\mbox{feature\_set} \gets \mathrm{FeatureSet()}$  \tcp{In \textsection{\ref{sec:design-train}}}
\setlength{\fboxsep}{2pt}\colorbox{cyan!12}{\parbox{\dimexpr\linewidth-2em-\algomargin}{%
$r_i \gets \min\!\big((1+r_t)/2,$\\
$\qquad 1-(1-r_t)\frac{\sum_{j=1}^{i}1/\sqrt{j}}{\sum_{j=1}^{K}1/\sqrt{j}}\big),\ \forall i \in [1,K]$\\
{\rmfamily// Per-rank target in \textsection\ref{sec:design-bh-target}}%
}}

\While{$N < K$}{
    \setlength{\fboxsep}{2pt}\colorbox{cyan!12}{\parbox{\dimexpr\linewidth-2em-\algomargin-\algomargin}{%
$\mbox{predicted\_avg\_recall} \gets$\\
$\qquad (\sum_{x=1}^{N} (r_x + \alpha(1-r_x))$\\
$\qquad\ + \sum_{r=N+1}^{K} T_{prob}(N, r)) / K$\\
{\rmfamily// Statistics-based forecast in \textsection\ref{sec:design-inference}}%
}}\;
    \uIf{$\mbox{predicted\_avg\_recall} \geq r_{t}$}{
        \textbf{break}\;
    }

    $\mbox{feature\_set}.\mathrm{mask}(\mbox{search\_set}[1..N])$\;
    \While{$\mathcal{M}.\mathrm{predict\_top\_1}(\mathrm{feature\_set}) < r_{N+1}$}{
        $\mbox{search\_set} \gets G.\mathrm{search\_multiple\_steps}(q, \mbox{search\_set})$\;
        $\mbox{feature\_set}.\mathrm{update}(\mbox{search\_set})$\;
    }
    
    $N \gets N + 1$\;
}
\KwRet{$\mbox{search\_set}[1..K]$}\;
\end{algorithm}

\subsection{Accelerating searches with statistics-based forecast}
\label{sec:design-inference}

\noindent
The basic search in Algorithm~\ref{alg:naive} invokes the learned model
at least once per rank,
so a top-$K$ query costs at least $K$ model calls, 
plus the iterations spent within each rank.
A known optimization, \emph{adaptive invocation frequency} from DARTH~\cite{darth},
is insufficient for top-$K$ queries.
Specifically, this optimization skips
some model calls \emph{within} a rank when searching for its top-$1$ vector (line 9--11),
but {\sys} still needs to invoke the model at least $K$ times per query.
Therefore, we propose a new \emph{statistics-based forecast} (line 5--7) that reduces invocations
\emph{across} ranks: once we are confident that the current search set already meets the recall
target, we skip the remaining ranks entirely.

\begin{figure}[!t]
        \begin{minipage}{0.99\linewidth}
       \centering    
        \includegraphics[width=\PaperLayout{0.99}{0.65}\linewidth, trim=0.25cm 23.0cm 21.5cm 0.25cm, clip]{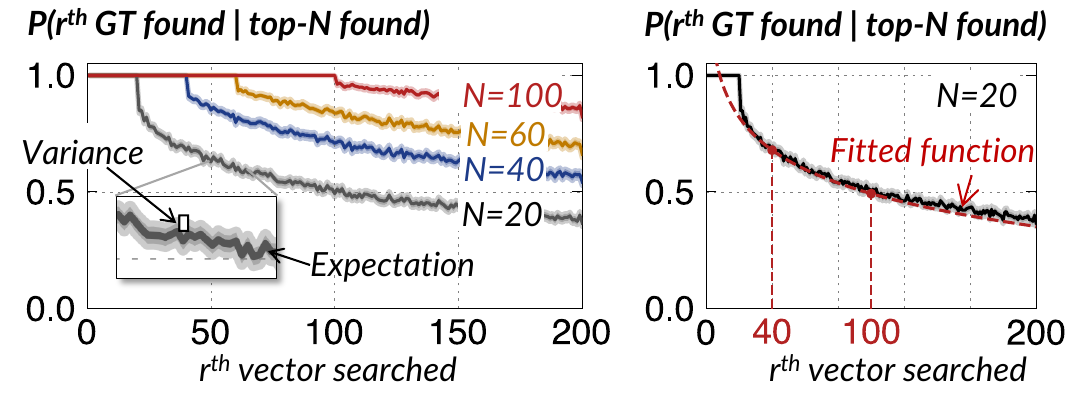}
        \end{minipage} \PaperLayout{\\[5pt]}{\\[-7pt]}
        \begin{minipage}{1\linewidth}
        \caption{\small{%
           (Left) An analysis of the conditional probability that the ground-truth $r^{th}$ closest vector 
           is in the search set, given $N$ collected nearest vectors.
            (Right) The conditional probability can be profiled via offline fitting using a few training-set samples. 
        }}
        \label{data:how-to-predict-average-recall}
        \end{minipage} \PaperLayout{\\[-15pt]}{\\[-8pt]}
        \end{figure} 

\stitle{Statistical forecast: insight and method. \,}
The key observation is that, once the search has already found the first $N$ ground-truth
neighbors ($N < K$), the probability that the $r$-th ($r > N$) ground-truth vector is
\emph{already} in the search set follows a clear and profileable pattern.
We denote this conditional probability by
$T_{prob}(N, r)=\Pr[r\text{-th groundtruth found}\allowbreak \mid \allowbreak \text{top-}N \text{ found}]$.
Empirically, for any fixed $r$, $T_{prob}(N, r)$ grows with $N$, as profiled in
{\fig{data:how-to-predict-average-recall}} (a): the more prefix vectors we have collected,
the more likely a further ground-truth neighbor has already been swept into the search set
as a by-product.
This lets us forecast the expected recall contribution of every rank $r > N$ directly from
$T_{prob}$, without invoking the learned model.
If the forecasted recall already meets the user-specified target $r_t$, we can skip the
learned search for all remaining ranks.

To realize this acceleration, 
we profile the index during training time to get a 2D conditional-probability table
that records the expectation of $T_{prob}(N, r)$.
The profiling is done over training queries,
and the time to build the above 2D lookup table is negligible compared to the training time of the learned model, 
because it only requires bookkeeping on the training set (see \textsection{\ref{sec:eval}} for more details).
We currently materialize the table up to size $200 \times 200$, as 200 is the maximum $K$ we observed in production workloads (see {\fig{data:darth-vs-omega-on-call}} (a)).
For unseen larger $K$ queries, we extrapolate the table along $r$ for a fixed $N$ by fitting a logarithmic decay function,
as shown in {\fig{data:how-to-predict-average-recall}} (b).

\stitle{Using the table during search. \,}
Consider a query with $K{=}200$. Suppose we have found $N{=}20$ ground-truth neighbors with the
iterative learned search, while the search set, which grows monotonically as the search
proceeds, already holds more than $K$ vectors.
Looking up {\fig{data:how-to-predict-average-recall}} (a), the
probability that the $200^{th}$ ground-truth is already in the search set is $0.3665$;
once $N$ grows to $40$, the probability rises to $0.5430$.
Applying the same lookup to every rank $r > N$ gives the forecasted recall@$K$:

\begin{equation}
    \label{eq:forecast}
\widehat{R}(N) \;=\; \frac{1}{K}\!\left(\,
\underbrace{\sum_{x=1}^{N}\!\bigl(r_x + \alpha(1{-}r_x)\bigr)}_{\text{calibrated found prefix}}
\;+\!\! \sum_{r=N+1}^{K} T_{prob}(N, r) \,\right).
\end{equation}

\noindent
The second sum is the expected recall contribution of the remaining ranks
(by linearity of expectation). For the first sum, we reuse the rank-wise
targets $r_x$ from \textsection\ref{sec:design-bh-target} as the marginal
correctness of each accepted rank, plus a small correction $\alpha(1{-}r_x)$
with $\alpha \in [0,1]$ that accounts for the search set continuing to
improve after rank $x$ is accepted; we set $\alpha$ close to $1$ in practice.
We stop the learned search as soon as $\widehat{R}(N) \geq r_t$, where $r_t$
is the user-specified recall target.
Note that although both the learned model and $T_{prob}$ are derived from the
aggregate training set, each decision is made from per-query context: every query
supplies its own search state---its own prefix length $N$ and its own accepted
ranks---so the stopping decision stays per-query, and different queries stop at
different $N$.\label{rev:per-query-stop}

\stitle{Put everything together: the optimized {\sys} search. \,}
Algorithm~\ref{alg:optimized-early-stop} shows the optimized {\sys} search
with the statistical forecast described in this section (as well as the rank-wise
confidence allocation in the previous section).
Unlike the basic search in Algorithm~\ref{alg:naive},
before using the model to find the next closest vector (line 10),
we leverage the profiled probability table $T_{prob}$ to forecast the expected mean recall
achieved so far (line 6).
If it exceeds the recall target (line 7),
we can stop the search early (line 8).

Besides the forecast, when using our model to specify the next closest vector (line 10--12),
we follow DARTH~\cite{darth} by adaptively calling the model
after searching multiple steps using the original graph index ($\mathrm{search\_multiple\_steps}(\cdot)$ (line 11)).
We omit a detailed description here for brevity.

\subsection{Generalizing to extremely large $K$s}
\label{sec:design-large-k}

\noindent
At very large $K$ (e.g., $K{=}500$),
{\sys}'s iterative refinement can still incur recall
loss from per-rank errors accumulated over many masked searches.
Though such extreme $K$s are rare in practice---our production trace
({\fig{data:darth-vs-omega-on-call}}\,(a)) has at most $K{=}200$---our
microbenchmark results reveal such degradation for $K > 200$
(\textsection\ref{sec:large-k-scalability}).
To generalize to extremely large $K$s,
we can let each model invocation certify a block of $K_{train}$ consecutive ranks,
i.e., replacing $\mathcal{M}.\mathrm{predict\_top\_1}$ in
Algorithm~\ref{alg:optimized-early-stop} with
$\mathcal{M}.\mathrm{predict\_top\_}K_{train}$ and cutting the number of masked
decisions from $K$ to $\lceil K/K_{train} \rceil$.
Because our trajectory-based features are rank-invariant
(\textsection\ref{sec:design-train}),
$K_{train}{>}1$ can be realized by retraining the base model at $K_{train}{>}1$.
Due to space constraints, 
Appendix~\ref{sec:appendix-k-train} describes the detailed extensions and 
\textsection\ref{sec:large-k-scalability} evaluates its effectiveness.

%% file: eval-v1.tex
\section{Evaluation}
\label{sec:eval}

\subsection{Evaluation Setup}
\label{sec:eval-setup}

\nospacestitle{System implementation. \,}
We implemented {\sys} in three codebases: hnswlib~\cite{hnswlib},
a widely used graph-based ANNS library,
{\company}'s internal vector search engine that implements the HNSW method, 
and Vamana~\cite{diskann}---the graph
index of DiskANN---for the generality study (\textsection\ref{sec:end2end-perf}).

\stitle{Testbed. \,}
Unless otherwise stated, all experiments run on a server with one Intel Xeon Gold 6430 CPU
(2.10\,GHz, 32 physical cores / 64 hyperthreads) and 1.5\,TB DRAM.

\begin{table}[!t]
\centering
\setlength{\abovecaptionskip}{2pt}
\setlength{\belowcaptionskip}{2pt}
\caption{Datasets used in our evaluation.}
\label{tab:datasets}
\setlength{\tabcolsep}{4pt}
\begin{tabular}{lllll}
\toprule
Dataset & Dim. & \#Vectors & Type & Metric \\ 
\midrule
BIGANN~\cite{bigann} & 128 & 100M--1B & uint8 & $L2$ \\ 
DEEP~\cite{deep} & 96 & 100M & float32 & $L2$    \\ 
GIST~\cite{gist} & 960 & 1M & float32 & $L2$     \\ 
Text2Image~\cite{text2image} & 200 & 1M & float32 & IP \\
MS MARCO~\cite{msmarco-web-search} & 768 & 1M & float32 & IP \\
Production 1 & 512 & 1.4M & int8 & $L2$  \\ 
Production 2 & 512 & 2.3M & int8 & $L2$  \\ 
Production 3 & 512 & 0.4M & int8 & $L2$ \\ 
\bottomrule
\end{tabular}
\vspace{-8pt}
\end{table}

\stitle{Datasets. \,}
We evaluate five open-source datasets
and three production collections from a {\company}
serving cluster that power key applications like image search and retrieval-augmented
generation (RAG). Table~\ref{tab:datasets} lists their details.
The open-source datasets cover Euclidean and inner-product search, with
dimensions from 96 to 960\footnote{We evaluate BIGANN at two scales:
BIGANN uses the first 100M vectors, while BIGANN-1B uses all 1B vectors.}.
Production datasets are run on {\company}'s internal engine; 
the open-source ones use hnswlib unless otherwise specified.

\begin{figure*}[!t]
    \centering
    \includegraphics[width=0.99\textwidth]{\PaperLayout{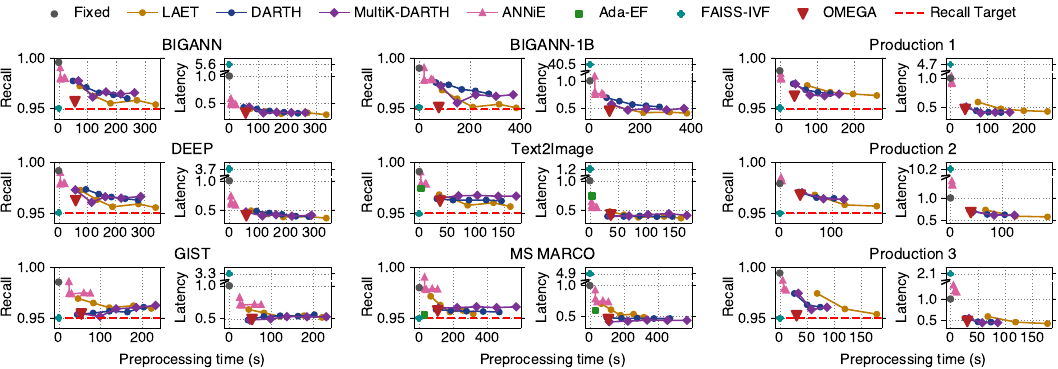}{data/eval-end2end-camera.pdf}}\PaperLayout{\\[5pt]}{\\[-5pt]}
    \begin{minipage}{1\linewidth}
    \caption{\small{
A comparison of recall and search latency across configurations under
different preprocessing time budgets. Search latency is
normalized to \textbf{Fixed}. 
        }}
    \label{fig:eval-end2end}
    \end{minipage} \PaperLayout{\\[-15pt]}{\\[-8pt]}
\end{figure*}

\stitle{Multi-$K$ traces. \,}
Because $K$ is application-dependent, we collected one-day traces for each
production dataset, recording both request embeddings and their $K$
values\footnote{Data were anonymized at collection;
the embeddings are derived representations that cannot be inverted to recover user data.}.
These traces give the $K$ distributions used in the evaluation.


\stitle{Training and testing queries. \,}
Following DARTH~\cite{darth}, we randomly sample training queries per dataset
(for open-source datasets, from their official learn sets).
For testing, production datasets use the corresponding traces directly;
open-source datasets use their standard query sets with $K$ drawn from the $K$ distributions of the production traces,
since they do not provide per-query $K$ values by default.
Training and testing queries are always disjoint:
open-source datasets follow the standard held-out split between the official
learn and query sets, so all results are reported on queries unseen during training.



\stitle{Baselines. \,}
We compare {\sys} against both learned ANNS search and conventional approaches: \\[-15pt]
\begin{itemize}[leftmargin=*,leftmargin=10pt,itemindent=0pt,itemsep=4pt]
    \item \textbf{Fixed} applies a fixed search step per $K$, 
    the default in hnswlib~\cite{hnswlib},
    Faiss~\cite{faiss}, and {\company}.
    Because {\company} hosts many multi-$K$ collections, it picks the step via a heuristic
    (larger $K$ gets a larger step); we adopt the same heuristic in our evaluations.

    \item \textbf{LAET}~\cite{laet} 
    is a state-of-the-art single-$K$ learned method:
    for queries at a given $K$, 
    it predicts the number of search steps needed to meet the recall target,
    adapting to per-query difficulty. 

    \item \textbf{DARTH}~\cite{darth} is another
    state-of-the-art learned method:
    during search, it uses runtime signals (e.g., the minimum observed distance)
    to predict whether the recall target is met.
    We optimized its implementation---via dynamic epoch adaptation
    ({\fig{data:training-loss}} left) and vectorized feature computation---reducing per-model training time
    to {4}\,\% of the open-source version at the same recall.

    \item \textbf{MultiK-DARTH}
    is a variant of DARTH that incorporates $K$ as a feature, so that a single
    model can accelerate queries with different $K$ values.

    \item \textbf{ANNiE}~\cite{annie}, like LAET,
    trains a separate model for each $K$ and predicts once per query the
    number of distance computations needed to meet a target recall.
    LAET makes this prediction after a fixed number of distance computations,
    whereas ANNiE chooses the prediction time adaptively and uses more
    features from the query's local graph.

    \item \textbf{Ada-ef}~\cite{ada-ef} estimates the HNSW \texttt{ef}
    hyperparameter by using statistics of the base-vector similarity distribution
    and distance statistics collected early in the graph search.
    Since Ada-ef does not support Euclidean ($L2$) distance, we report
    it on Text2Image and MS MARCO, the two inner-product datasets.

    \item \textbf{FAISS-IVF}~\cite{faiss} partitions vectors into clusters
    and searches only selected clusters, using \texttt{nprobe} to control the
    recall--latency trade-off.
    We tune \texttt{nprobe} to the 0.95 recall target for each dataset.
\end{itemize}

\subsection{Overall recall and latency with different preprocessing time budgets}
\label{sec:end2end-perf}

\nospacestitle{Setup. \,}
{\fig{fig:eval-end2end}} compares the mean recall and latency across nine
datasets under varying preprocessing time budgets.
Note that we have excluded the static index/dataset feature computation time of
ANNiE and Ada-ef, which is non-negligible on large datasets and makes the graph hard to interpret.
Because Fixed is {1.4--3.4\,$\times$} slower than LAET, DARTH, and {\sys},
we report latencies normalized to it for clarity.
Reported numbers are means over multi-$K$ queries. 

\begin{figure*}[!t]
    \centering
    \includegraphics[width=0.99\textwidth]{\PaperLayout{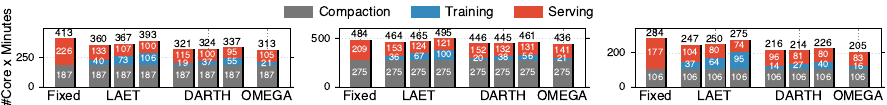}{data/eval-cpu-camera.pdf}}\PaperLayout{\\[7pt]}{\\[-5pt]}
    \begin{minipage}{1\linewidth}
    \caption{\small{
A comparison of CPU time usage on production datasets
(Production 1--3, \PaperLayout{left to right}{top left, top right, and bottom left}).
For clarity, LAET and DARTH are shown as grouped bars that summarize the
setups with 1, 2, and 3 trained models for accelerating frequently accessed $K$ values.
        }}
    \label{fig:eval-cpu}
    \end{minipage} \PaperLayout{\\[-15pt]}{\\[-8pt]}
\end{figure*}

\stitle{Preprocessing time. \,}
{\sys} trains a single $K{=}1$ model per dataset,
so its preprocessing time is constant.
The preprocessing time of LAET and DARTH scales with the number of trained $K$ models,
reaching up to {6.6}\,$\times$ that of {\sys}.
MultiK-DARTH trains a single model, but it still needs enough training data per $K$
to converge, so its preprocessing time also scales with the number of covered $K$s.
For a single model, {\sys}'s preprocessing time is comparable to that of DARTH
({0.63--1.22}\,$\times$ across datasets).

\stitle{Recall results. \,}
{\sys} meets the 95\% recall target in every configuration---a high bar
adopted by prior work~\cite{DBLP:conf/osdi/MohoneySTCPIRV25,darth,laet,rabitq,milvus,bang}.
DARTH and LAET also generally exceed this target
because they prioritize models
for the most frequently accessed $K$s and serve an unseen $K$ using the closest
trained model. In many datasets, the most frequent $K$s are large (e.g.,
$K{=}100$ in Production 1 and 2), so reusing these models for smaller-$K$
queries causes oversearch that raises recall.
This also explains why the recall of DARTH and LAET can
decrease as preprocessing time increases (e.g., on BIGANN): newly added
smaller-$K$ models search less than the reused large-$K$ model,
which lowers latency.

\stitle{Latency results. \,}
Against LAET, DARTH, MultiK-DARTH, ANNiE, and Ada-ef at their
lowest-preprocessing configurations, {\sys} reduces mean latency by up to 65\%
(average: 23\%; median: 20\%) across all 38 applicable dataset--baseline comparisons.
Specifically, {\sys} reduces
mean latency by 7--36\% versus DARTH, 3--25\% versus MultiK-DARTH, and 8--21\%
versus LAET on BIGANN, BIGANN-1B, DEEP, and the three production workloads.
With this small preprocessing budget, DARTH and LAET reuse models trained for
frequent, often large $K$s and oversearch smaller-$K$ queries; adding more
$K$-specific models reduces this oversearch but increases preprocessing time.
On the other three datasets---GIST, Text2Image, and MS MARCO---{\sys} remains
within 9\% of DARTH and MultiK-DARTH while reducing mean latency by 13--37\%
versus LAET.
Among the other adaptive baselines, {\sys} reduces mean latency by 23--60\%
relative to ANNiE's lowest-latency configuration across all nine datasets,
and by 44\% and 25\% relative to Ada-ef on Text2Image and MS MARCO, respectively.
Finally, the cluster-based FAISS-IVF meets the recall target, but its mean latency is
1.2--40.5\,$\times$ that of Fixed across all configurations.

\begin{figure*}[!t]
    \centering
    \includegraphics[width=0.99\textwidth]{\PaperLayout{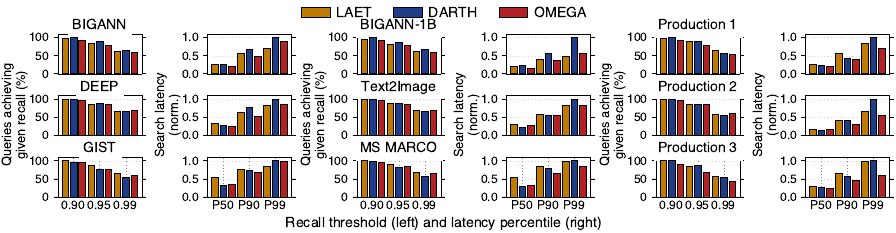}{data/eval-percent-camera.pdf}}\PaperLayout{\\[5pt]}{\\[-8pt]}
    \begin{minipage}{1\linewidth}
    \caption{\small{
A comparison of P50, P90, and P99 recall/latency across datasets for all methods using the same preprocessing time budget (training one model).
Latency is normalized to \textbf{Fixed} for clarity.
        }}
    \label{fig:eval-percent}
    \end{minipage} \PaperLayout{\\[-15pt]}{\\[-8pt]}
\end{figure*}

\begin{figure}[!t]
    \centering
    \includegraphics[width=\PaperLayout{0.99}{0.65}\linewidth]{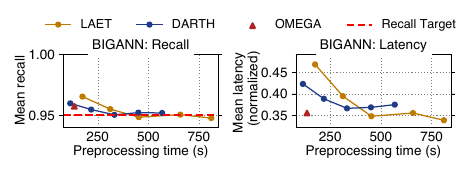}\PaperLayout{\\[0pt]}{\\[-13pt]}
    \begin{minipage}{1\linewidth}
    \caption{\small{
Recall and search latency on Vamana~\protect\cite{diskann} (BIGANN),
following the same setup as {\fig{fig:eval-end2end}}.
        }}
    \label{fig:eval-end2end-diskann-main}
    \end{minipage} \PaperLayout{\\[-15pt]}{\\[-8pt]}
\end{figure}

\stitle{Generality: results on Vamana (DiskANN~\cite{diskann}). \,}
We further evaluate {\sys} on Vamana, another popular graph index.
{\fig{fig:eval-end2end-diskann-main}} reports the results on BIGANN due to space limitations,
while 
Appendix~\ref{sec:appendix-features} reports all results on other 
open-source datasets. 
{\sys} meets the recall target with a single model while reducing search
latency by 16\,\% versus DARTH and 24\,\% versus LAET at a comparable
preprocessing budget.
{\sys} generalizes because the trajectory feature is also effective on Vamana,
as shown in the appendix.

\stitle{Total CPU time used. \,}
Though learned methods like {\sys} and DARTH incur extra CPU time,
the savings in online serving CPU time outweigh this cost.
{\fig{fig:eval-cpu}} reports the CPU time provisioned per collection over a one-day trace
in which each collection triggers a compaction. We derive the serving time by replaying
the trace and aggregating the per-query search times.
{\sys} cuts total CPU time by {10--28}\,\% vs Fixed, {6--25}\,\% vs LAET, and {2--9}\,\% vs DARTH.
The low preprocessing cost of Fixed cannot offset its high serving cost, which comes from its long search latency.
On the other hand, LAET and DARTH have a preprocessing--serving trade-off,
which {\sys} avoids.

\begin{figure*}[!t]
    \centering
    \includegraphics[width=\PaperLayout{0.95}{0.99}\textwidth]{\PaperLayout{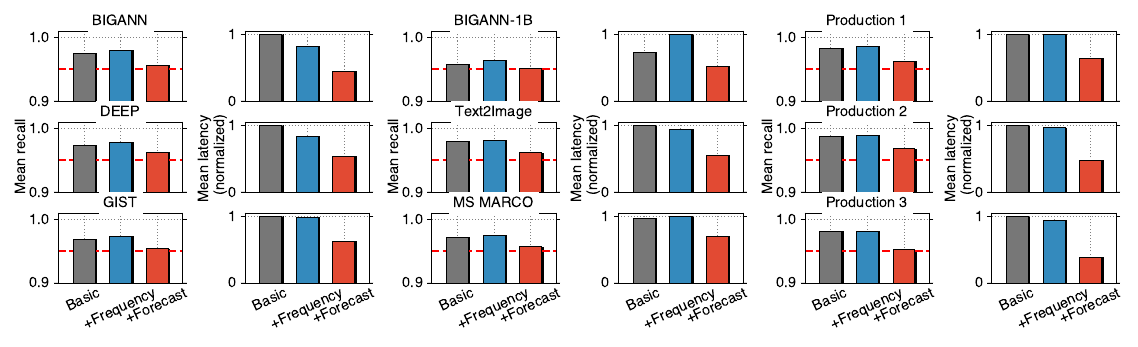}{data/ablation-camera.pdf}}\PaperLayout{\\[-5pt]}{\\[-13pt]}
    \begin{minipage}{1\linewidth}
    \caption{\small{
        Ablation study of {\sys}'s adaptive invocation frequency and statistical forecast optimizations. 
        }}
    \label{fig:ablation}
    \end{minipage} \PaperLayout{\\[-10pt]}{\\[-8pt]}
\end{figure*}

\subsection{Results on median and tail}
\label{sec:per-query-performance}

\noindent
\nospacestitle{Setup. \,}
{\fig{fig:eval-percent}} reports the fraction of queries passing recall thresholds $\{0.90, 0.95, 0.99\}$ and the P50, P90, and P99 search latency
on each dataset. 
The setup is the same as the previous section,
and all systems adopt the same preprocessing cost:
{\sys} trains at $K=1$, and other learned baselines train
on the most frequently accessed $K$ in each dataset.

\stitle{Recall results. \,}
Across all nine datasets, {\sys} achieves 0.90 recall on {88--93}\,\% of
queries, within {9}\,pp of the better of LAET and DARTH.
Its per-query recall remains comparable at the stricter 0.95 and 0.99
thresholds, where the maximum gaps are {17}\,pp and {13}\,pp, respectively.

\stitle{Latency results. \,} 
{\sys} generally achieves the lowest latency across P50, P90, and P99 while
maintaining comparable per-query recall.
In particular, it is consistently faster than DARTH at P90 and P99 across all
nine datasets, reducing latency by {1--35}\,\% and {3--44}\,\%, respectively.

\subsection{Ablation Study}
\label{sec:ablation-study}

\begin{figure}[!t]
        \begin{minipage}{1\linewidth}
        \centering
         \includegraphics[width=\PaperLayout{\linewidth}{0.65\linewidth}]{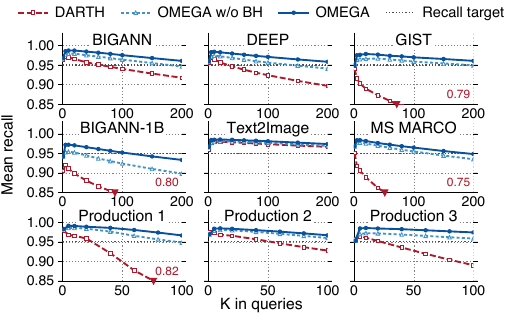}
        \end{minipage} \PaperLayout{\\[6pt]}{\\[-5pt]}
        \begin{minipage}{1\linewidth}
        \caption{\small{%
            Ablation study of {\sys}'s trajectory-based features and BH rank-wise target allocation.
            $\blacktriangledown$ marks a curve leaving the plotted range, labeled with its recall at the largest $K$.
        }}
        \label{fig:ablation-generalizability}
        \end{minipage} \PaperLayout{\\[-10pt]}{\\[-8pt]}
        \end{figure}

\noindent
\nospacestitle{Effectiveness of trajectory-based features (\textsection{\ref{sec:design-train}}). \,}
{\fig{fig:ablation-generalizability}} compares DARTH and {\sys} w/o BH---{\sys}
without the Benjamini--Hochberg (BH) rank-wise target allocation of
\textsection{\ref{sec:design-robustness}}---both trained on $K{=}1$ and
generalized to multi-$K$ queries via Algorithm~\ref{alg:naive}.
DARTH uses absolute-distance features and {\sys} uses trajectory-based features.
Across all nine datasets, {\sys} w/o BH consistently outperforms DARTH as $K$
increases.  At $K{=}200$, for example, DARTH degrades to 0.75 on MS MARCO and
0.80 on BIGANN-1B, whereas {\sys} w/o BH retains 0.94 and 0.90 recall.
The gap confirms that trajectory-based features are essential for generalizing to multiple $K$s.
We omit LAET because it cannot support the iterative generalization.

\stitle{Effectiveness of rank-wise target allocation (\textsection{\ref{sec:design-robustness}}). \,}
{\fig{fig:ablation-generalizability}} further compares {\sys} w/o BH and {\sys}. 
{\sys} w/o BH works on most datasets, but can still fall short of the 0.95 recall target as $K$ increases.
BH improves recall on every dataset and, at the largest plotted $K$, meets the
target on seven of nine datasets while raising MS MARCO and BIGANN-1B to 0.948
and 0.933, respectively.

\stitle{Effectiveness of adaptive prediction and statistics-based forecasting (\textsection{\ref{sec:design-inference}}). \,}
As shown in {\fig{fig:ablation}}, the baseline invokes the model at a fixed interval of 50 steps.
Incorporating the adaptive invocation frequency from DARTH~\cite{darth} (+Frequency)
has dataset-dependent benefits because it cannot reduce multi-$K$ invocations.
Adding our statistics-based forecast (+Forecast) achieves the lowest latency
while meeting the recall target on all nine datasets: it reduces mean latency
by {27--61}\,\% over the baseline and by {30--59}\,\% over +Frequency.

\begin{figure}[!t]
        \begin{minipage}{1\linewidth}
        \centering
         \includegraphics[width=\PaperLayout{0.99}{0.65}\linewidth]{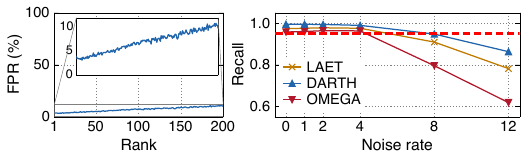}
        \end{minipage} \PaperLayout{\\[5pt]}{\\[-8pt]}
        \begin{minipage}{1\linewidth}
        \caption{\small{
            (a) Rank-wise false positive rate of {\sys} across masked ranks;
            the inset magnifies the 0--12\% band.
            (b) Recall of LAET, DARTH, and {\sys} 
            with increased train--test distribution shifts. 
        }}
        \label{fig:revision-robustness}
        \end{minipage} \PaperLayout{\\[-15pt]}{\\[-8pt]}
\end{figure}

\stitle{Robustness of {\sys}. \,}
First, we measure the
false positive rate (FPR), i.e., the probability that the model accepts a rank as
found when the corresponding ground-truth neighbor has not been found.
As shown in {\fig{fig:revision-robustness}}~(a), the rank-wise FPR grows 
slowly as the rank increases: from 3.9\% at rank 1 to 9.9\% at rank 200,
with an overall FPR of 5.3\%.

Second, we evaluate whether {\sys} is robust when the
distribution of the test queries deviates from that of the training queries.
In {\fig{fig:revision-robustness}} (b),
{\sys} remains robust under moderate noise ($\le$4\%),
while under larger noise all learned methods lose recall.

\begin{figure*}[!t]
    \centering
    \includegraphics[width=0.99\textwidth]{\PaperLayout{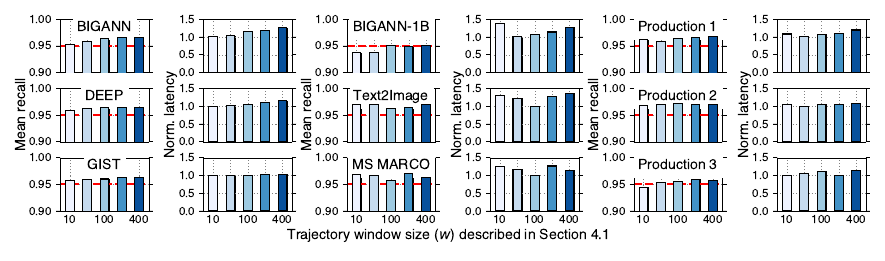}{data/sensitivity-v1-camera.pdf}}\PaperLayout{\\[-7pt]}{\\[-12pt]}
    \begin{minipage}{1\linewidth}
    \caption{\small{
        Sensitivity of {\sys} to the trajectory window size $w$.
        Latency is normalized to the fastest $w$ of each dataset.
        }}
    \label{fig:sensitivity}
    \end{minipage} \PaperLayout{\\[-10pt]}{\\[-8pt]}
\end{figure*}

\subsection{Sensitivity to trajectory window size}
\label{sec:sensitivity-to-hyperparameters}

\noindent
{\sys} exposes only one hyperparameter: the trajectory window size $w$
(\textsection{\ref{sec:design-train}}).
All experiments except the one in this section use $w{=}100$.
As shown in {\fig{fig:sensitivity}},
{\sys} is robust to $w$ across all nine datasets.
Across all datasets, $w{=}100$ meets the recall target with near-minimum
latency; on BIGANN-1B, Text2Image, and MS MARCO, it is the lowest-latency
configuration that meets the target.
This is because search trajectories are bounded in length in practice,
so a sufficiently large window covers the entire trajectory.

\subsection{Performance on extremely large $K$s}
\label{sec:large-k-scalability}

\begin{figure}[!t]
        \begin{minipage}{1\linewidth}
        \centering
         \includegraphics[width=\PaperLayout{0.99}{0.70}\linewidth]{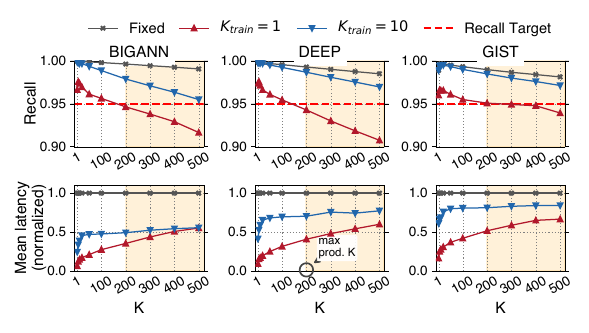}
        \end{minipage} \PaperLayout{\\[0pt]}{\\[-12pt]}
        \begin{minipage}{1\linewidth}
        \caption{\small{%
            (Top) recall and (bottom) normalized mean latency of {\sys} as $K$ grows.
            The orange region marks $K{>}200$, beyond the maximum $K$ observed in
            our production traces.
        }}
        \label{fig:large-k-overhead}
        \end{minipage} \PaperLayout{\\[-10pt]}{\\[-8pt]}
\end{figure}

\noindent
{\sys} meets the recall target at every $K$ seen in production.
For larger $K$s, changing its base model to top-$10$ keeps recall at or near the
target up to $K{=}500$, at the cost of higher latency at small $K$s.
As $K$ grows, the iterative refinement in {\sys} can lose recall from
accumulated per-iteration errors and can add latency.
{\fig{fig:large-k-overhead}} shows the results
on BIGANN, DEEP, and GIST for $K$s far larger than those observed
in our production traces ($K{=}200$).
For the $K$s seen in production, {\sys} reaches the recall target.
For $K>200$, its recall drops
to 0.907--0.949 across the three datasets due to
accumulated per-iteration errors.
Changing the base model to top-$10$ removes this degradation
(either by retraining at $K_{train}{=}10$, or by reusing the existing top-$1$ model with a larger block size; see Appendix~\ref{sec:appendix-k-train}):
with $K_{train}{=}10$, {\sys} stays near or above the 0.95 target up to
$K{=}500$ on all three datasets, and reaches 0.955--0.981.
Using a top-$10$ base model may increase latency, especially at small $K$s,
due to oversearch: at $K{\le}10$, its latency is
{2.5--4.4}$\times$ that of top-$1$ across the three datasets.
Even so, {\sys} remains faster than Fixed at all $K$s---its normalized
latency stays within 0.524--0.841 for $K>200$ and is 0.556--0.841 at $K{=}500$.

%% file: relatedwork.tex
\section{Related Work}
\label{sec:related}

\nospacestitle{Learned search for vector databases. \,} 
LAET~\cite{laet} and ANNiE~\cite{annie} predict per-query search budgets
for a target $K$, while DARTH~\cite{darth} predicts when to stop from the search state.
These methods train for specific $K$ values; supporting multi-$K$ workloads
with separate models increases preprocessing cost (\textsection{\ref{sec:overview}}).
A recent method, Ada-ef~\cite{ada-ef}, estimates the HNSW \texttt{ef}
hyperparameter by using base-vector similarity-distribution statistics and
distance statistics collected early in the graph search; it handles multi-$K$
with trivial preprocessing, but its theory applies only to inner-product/cosine
distance and does not yet extend to Euclidean ($L2$)---the metric used by all
our production traces.
%

\stitle{Graph-based ANN search optimizations. \,}
Besides learned methods to accelerate graph-based ANN search,
recent works also accelerate the computation mechanism of graph search,
including using approximated distance computation to accelerate each expansion~\cite{ADSampling,BetterADSampling,FINGER,DBLP:journals/pvldb/DengCZWZZ24},
or selecting a better entry point to shorten the search path~\cite{DBLP:conf/mir/HezelBSJ24,LSH-APG,DBLP:conf/iclr/SunGK23}.
They are orthogonal to {\sys} since they do not change the graph traversal patterns.
Meanwhile, several works optimize graph searches on slow storage (e.g., SSD)~\cite{diskann,Starling,pipelinessd}.
As these works also do not modify the search trajectory and the graph structure, {\sys} can be combined with them as well.

\stitle{Cluster-based ANN search. \,}
Besides graph-based ANN search, partition/cluster-based indexes are widely
used alternatives~\cite{spann,DBLP:conf/osdi/MohoneySTCPIRV25,spfresh,LIRA}.
{\sys} currently targets graph traversal because graph indexes deliver
superior performance in many workloads,
are common in open-source vector databases~\cite{qdrant-indexing,weaviate-vector-index},
and are the dominant choice in {\company}.
We also compare against tuned cluster-based baselines in {\fig{fig:eval-end2end}}.
Extending {\sys} to cluster probing is orthogonal and left as future work.\label{rev:cluster-anns}

%% file: conclusion.tex
\section{Conclusion}
\label{sec:conclusion}

\noindent
This paper shows that learned search can support arbitrary $K$
with only a single top-$1$ model.
We do so by reducing top-$K$ to repeated masked top-$1$ refinement,
and propose three techniques for this reduction:
trajectory features that generalize across ranks,
a rank-wise confidence allocation scheme that constrains generalization error,
and a statistical forecast for acceleration.
Across nine datasets, both open-source and production,
{\sys} meets the 0.95 recall target with one $K$-independent model
and achieves a better online-offline tradeoff than the state-of-the-art
learned and traditional methods.

%% file: ack.tex
\noindent
We sincerely thank the SIGMOD reviewers for their insightful comments
and constructive suggestions.
We also thank Shitong Chen for helping us prepare the evaluation environment.
This work was supported in part by
the National Natural Science Foundation of China (No.~62572302, 62272291)
and the Alibaba Innovative Research Program.

%% file: appendix.tex
\section{Why GPU Acceleration has Limited Impact on Training Speed}
\label{sec:appendix-gpu}

\subsection{Limited GPU Acceleration Potential for LightGBM-based GBDT}
\label{sec:appendix-lightgbm}

\noindent
We found that GPU provides limited speedup for training popular models used by state-of-the-art learned ANNS methods including DARTH~\cite{darth} and LAET~\cite{laet}. 
This limited speedup stems from LightGBM's fundamental tree construction strategy: 
it utilizes a leaf-wise growth approach where trees grow node by node, 
creating sequential dependencies between tree nodes that hinder parallel execution on GPU architectures~\cite{DBLP:conf/nips/KeMFWCMYL17}. 
While LightGBM does support GPU acceleration for histogram construction (a preprocessing step before tree building), 
the computational complexity of this operation scales linearly with input feature dimensionality~\cite{DBLP:conf/nips/KeMFWCMYL17}. 
Because learned search typically operates on low-dimensional feature spaces (e.g., 11 dimensions), as higher-dimensional features would be too slow for online prediction,
the histogram construction time becomes negligible compared to the overall tree building process. 
Consequently, in our scenario, the speedup from GPU acceleration is limited.

\subsection{Why Choose LightGBM over XGBoost}
\label{sec:appendix-xgboost}

\noindent
Although several models, e.g., XGBoost, employ a level-wise (depth-wise) tree growth strategy that expands trees layer by layer and are more amenable to GPU acceleration~\cite{xgboost}, 
{\sys} and previous learned ANNS works~\cite{laet,darth} selected LightGBM due to its superior inference performance. 
Specifically, LightGBM demonstrates exceptional performance for single-input predictions, 
achieving average inference times of $<10\mu$s on a single CPU core for our 11-dimensional feature space. 
This efficiency is critical for our application context: 
vector searches on in-memory graph indexes already achieve extremely low latency (average exploration time $<1\mu$s per point). 
To ensure that the performance gains from the early stopping mechanism outweigh the prediction overhead, 
we must minimize the computational cost of each individual prediction.

\section{Generality of Our Trajectory-based Features}
\label{sec:appendix-features}

\begin{figure}[!t]
        \PaperLayout{\hspace{-3mm}}{}
        \begin{minipage}{1\linewidth}
        \centering    
        \includegraphics[width=\PaperLayout{0.99}{0.7}\linewidth, trim=0.25cm 20cm 12.6cm 0.25cm, clip]{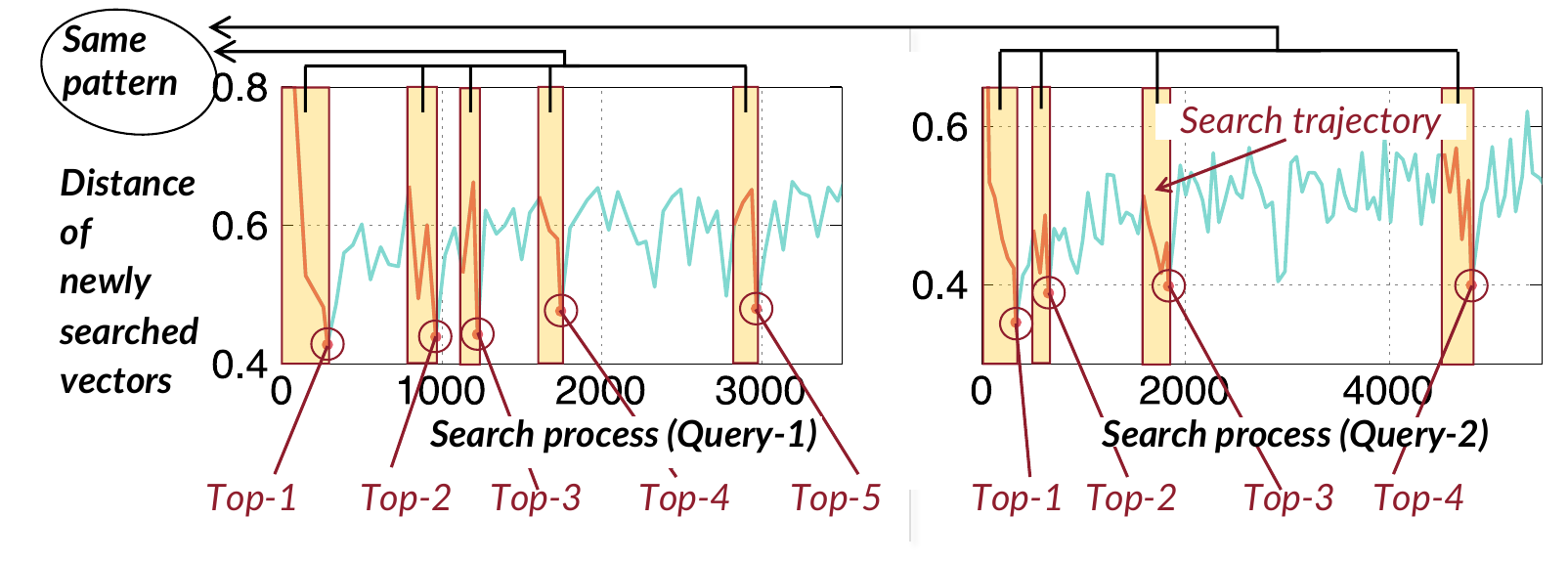}
        \end{minipage} \PaperLayout{\\[5pt]}{\\[-7pt]}
        \begin{minipage}{1\linewidth}
        \caption{\small{%
    Distance trajectory patterns observed on Vamana~\protect\cite{diskann} for two different queries on a production
    collection in {\company}.
        }}
        \label{fig:search-trajectory-breakdown-diskann}
        \end{minipage} \PaperLayout{\\[-10pt]}{\\[-8pt]}
        \end{figure} 

\noindent
{\fig{fig:search-trajectory-breakdown-diskann}} illustrates the distance trajectory patterns observed
on Vamana~\cite{diskann}---another popular graph index structure for vector databases on disks.
We can see that it also follows our observations described in \textsection{\ref{sec:design-train}},
so our learned search is also applicable to Vamana.

\begin{figure}[!t]
    \centering
    \includegraphics[width=\PaperLayout{0.99}{0.60}\linewidth]{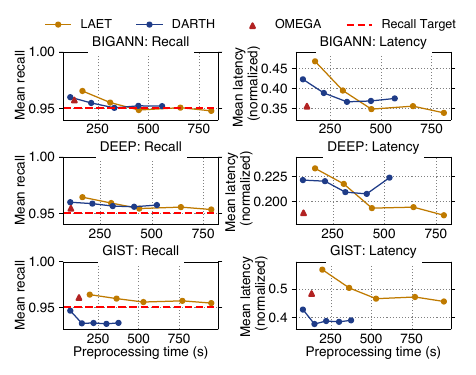}\PaperLayout{\\[0pt]}{\\[-12pt]}
    \begin{minipage}{1\linewidth}
    \caption{\small{
A comparison of the recall and search latency with different preprocessing time budgets across the three open-source datasets on Vamana~\protect\cite{diskann}.
The search latency is normalized to the \textbf{Fixed} baseline, which sets a large, conservative search step for all queries.
        }}
    \label{fig:eval-end2end-diskann}
    \end{minipage} \PaperLayout{\\[-10pt]}{\\[-8pt]}
\end{figure}

{\fig{fig:eval-end2end-diskann}} further reports the end-to-end recall and
search latency of {\sys} and the baselines on Vamana across the three open-source datasets,
following the same setup as {\fig{fig:eval-end2end}}.
The results show that {\sys} retains its advantage over LAET and DARTH on Vamana:
it meets the 0.95 recall target on all datasets while achieving the lowest
search latency at a comparable preprocessing budget,
confirming that our learned search generalizes beyond hnswlib.
The only exception is GIST, where {\sys}'s latency is higher than DARTH's;
this is because DARTH undersearches on GIST and therefore fails to meet the
recall target, so its lower latency does not reflect a fair comparison.

\clearpage
\section{QPS-recall Curve Compared to Tuned Fixed}
\label{sec:appendix-qps-recall}

\begin{figure}[!t]
    \centering
    \includegraphics[width=\PaperLayout{0.99}{0.65}\linewidth]{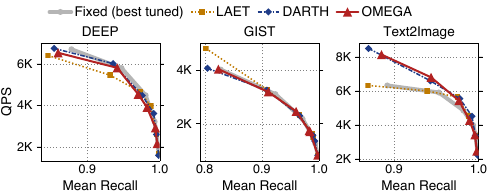}\PaperLayout{\\[5pt]}{\\[-6pt]}
    \begin{minipage}{1\linewidth}
    \caption{\small{QPS--recall results.}}
    \label{fig:qps-recall-frontier}
    \end{minipage} \PaperLayout{\\[-10pt]}{\\[-8pt]}
\end{figure}

\noindent
One may wonder whether {\sys} can outperform baselines with tuned best parameters 
of the Fixed baseline described in \textsection{\ref{sec:eval-setup}}. 
This section reports the query-per-second vs. recall curve
on different datasets.
Fixed sweeps efSearch from 100 to 2000:
each point uses the best single parameter at its recall level.
The learned methods run under the same efSearch bounds with
increasing recall targets. 
{\fig{fig:qps-recall-frontier}} presents the results. 
All experiments use $K{=}100$ and the same testbed described in \textsection{\ref{sec:eval-setup}}.
DARTH and LAET adopt the model trained for $K{=}100$ queries,
while {\sys} directly uses a model trained for $K{=}1$ with our optimized search.

Tuning improves Fixed, but one hyperparameter still
serves all queries: to meet a recall target, it must be sized for the hard
queries, so the easy queries oversearch.
The learned methods (LAET, DARTH, and {\sys}) instead terminate each query
individually, reducing the number of vectors searched by up to 37\%
compared to the tuned Fixed under matched settings.
The advantage is most visible on Text2Image:
{\sys} reaches 8.1\,K QPS at 0.89 mean recall,
1.29$\times$ that of the tuned Fixed at a similar recall.

\begin{algorithm*}[t]
\caption{Optimized {\sys} Search with $K_{train}>1$}
\label{alg:optimized-early-stop-k-train}
\DontPrintSemicolon
\KwIn{Graph index $G$, query vector $q$, recall target $r_{t}$, number of results $K$, training block size $K_{train}$, {\sys} model $\mathcal{M}$, profiled probability table $T_{prob}$}
\KwOut{$K$ nearest vectors of $q$}

$N \gets 0$ \tcp{Number of certified ranks}
$\mbox{search\_set} \gets \mathrm{PriorityQueue}()$\;
$\mbox{feature\_set} \gets \mathrm{FeatureSet()}$\;
\setlength{\fboxsep}{2pt}\colorbox{cyan!12}{\parbox{\dimexpr\linewidth-2em-\algomargin}{%
$r_i \gets \min\!\left((1+r_t)/2,\ 1-(1-r_t)\frac{\displaystyle\sum_{j=1}^{\lceil i / K_{train} \rceil}1/\sqrt{j}}{\displaystyle\sum_{j=1}^{\lceil K / K_{train} \rceil}1/\sqrt{j}}\right),\ \forall i \in [1,K]$ {\rmfamily// Block-wise BH target}%
}}

\While{$N < K$}{
    \setlength{\fboxsep}{2pt}\colorbox{cyan!12}{$N_{next} \gets \min(N+K_{train}, K)$}\;
    $\mbox{predicted\_avg\_recall} \gets (\sum_{x=1}^{N} (r_x + \alpha(1-r_x)) + \sum_{r=N+1}^{K} T_{prob}(N, r)) / K$ \tcp{Statistics-based forecast in \textsection\ref{sec:design-inference}}
    \uIf{$\mbox{predicted\_avg\_recall} \geq r_{t}$}{
        \textbf{break}\;
    }

    $\mbox{feature\_set}.\mathrm{mask}(\mbox{search\_set}[1..N])$\;
    \setlength{\fboxsep}{2pt}
    \While{\colorbox{cyan!12}{$\mathcal{M}.\mathrm{predict\_top\_k}(\mathrm{feature\_set}, K_{train}) < r_{N_{next}}$}}{
        $\mbox{search\_set} \gets G.\mathrm{search\_multiple\_steps}(q, \mbox{search\_set})$\;
        $\mbox{feature\_set}.\mathrm{update}(\mbox{search\_set})$\;
    }

    \setlength{\fboxsep}{2pt}\colorbox{cyan!12}{$N \gets N_{next}$}\;
}
\KwRet{$\mbox{search\_set}[1..K]$}\;
\end{algorithm*}

Exploring fewer points with learned methods,
however, does not always improve the QPS. 
This is because the prediction overhead may offset the reduced searches,
and for workloads where a fixed hyperparameter is sufficient for most test queries,
the learned methods can achieve similar or worse results. 
On DEEP, at similar recall, the QPS of all methods is close;
at the low-recall end, Fixed is even slightly ahead:
6.7\,K QPS at 0.88 mean recall, vs. 6.5\,K at 0.86 for {\sys}.
On GIST, the learned methods achieve similar QPS to the tuned Fixed
at similar recall, and at the highest recall level (0.998),
the QPS of all four methods is within 2\%.

Note that tuned Fixed is typically impractical to deploy:
selecting the best parameters requires a
per-dataset, per-$K$, per-recall-target tuning, repeated after every
compaction.
As a result, hnswlib, Faiss, and our production deployment default to
heuristic search steps, so the main evaluation
(\textsection{\ref{sec:eval}}) uses this heuristic Fixed.
Existing learned methods like DARTH automate the per-recall-target tuning,
while {\sys} further avoids the per-$K$ tuning.

\section{Details on generalizing {\sys} to extremely large $K$s}
\label{sec:appendix-k-train}

\noindent
\textsection{\ref{sec:design-large-k}} describes the necessity
of training a base model beyond top-1 when the workload has an
extremely large $K$.
Specifically, we can instead train the base model with $K_{train}{>}1$ and
let each model invocation early-stop a block of $K_{train}$ consecutive ranks.
All our designs---including the trajectory-based features,
the rank-wise confidence allocation scheme, and
the statistical forecast for acceleration---still apply.
The only change is that Algorithm~\ref{alg:optimized-early-stop}
needs to be refined for block-wise execution.

Algorithm~\ref{alg:optimized-early-stop-k-train} shows the generalized search
with a model capable of deciding whether the current search set contains the
next $K_{train}$ target vectors.
The main difference is that the outer loop first computes
$N_{next}=\min(N+K_{train}, K)$ and then uses
$\mathcal{M}.\mathrm{predict\_top\_k}(\mathrm{feature\_set}, K_{train})$ to certify
all ranks up to $N_{next}$ at once.